\documentclass{article}

\usepackage{arxiv}

\usepackage[utf8]{inputenc} 
\usepackage[T1]{fontenc}    
\usepackage{hyperref}       
\usepackage{url}            
\usepackage{booktabs}       
\usepackage{amsfonts}       
\usepackage{nicefrac}       
\usepackage{microtype}      
\usepackage{cleveref}       
\usepackage{tablefootnote}
\usepackage[table]{xcolor}

\usepackage{placeins}
\usepackage{graphicx}
\graphicspath{ {./images/} }

\usepackage{natbib}
\usepackage{doi}

\usepackage{numprint}
\npthousandsep{,}

\usepackage{caption} 

\title{Thousands of AI Authors on the Future of AI}

\date{January 2024}

\author{
        Katja Grace\thanks{Corresponding author}\thanks{Equal Contribution}\thanks{Shared senior authorship} \\
	AI Impacts\\
	Berkeley, California \\
        United States \\
	\texttt{katja@aiimpacts.org} \\
 \And
        Harlan Stewart\footnotemark[2]\\
        AI Impacts\\
        Berkeley, California\\
        United States\\
\And
	Julia Fabienne Sandkühler\footnotemark[2] \\
	Department of Psychology\\
	University of Bonn\\
	Germany \\
\And
        Stephen Thomas\footnotemark[2]\\
        AI Impacts\\
        Berkeley, California\\
        United States\\
\And
        Ben Weinstein-Raun \\
        Independent \\
        Berkeley, California \\
        United States \\
\And
        Jan Brauner\footnotemark[3] \\
        Department of Computer Science \\
        University of Oxford \\
        United Kingdom \\
\And
        Richard C. Korzekwa\footnotemark[2]\footnotemark[3]\\
        AI Impacts\\
        Berkeley, California\\
        United States\\
}

\begin{document}
\maketitle

\setcounter{footnote}{0} 

\begin{abstract}
In the largest survey of its kind, we surveyed \numprint{2778} researchers who had published in top-tier artificial intelligence (AI) venues, asking for their predictions on the pace of AI progress and the nature and impacts of advanced AI systems. The aggregate forecasts give at least a 50\% chance of AI systems achieving several milestones by 2028, including autonomously constructing a payment processing site from scratch, creating a song indistinguishable from a new song by a popular musician, and autonomously downloading and fine-tuning a large language model. If science continues undisrupted, the chance of unaided machines outperforming humans in every possible task was estimated at 10\% by 2027, and 50\% by 2047. The latter estimate is 13 years earlier than that reached in a similar survey we conducted only one year earlier \citep{grace2022}. However, the chance of all human occupations becoming fully automatable was forecast to reach 10\% by 2037, and 50\% as late as 2116 (compared to 2164 in the 2022 survey).

Most respondents expressed substantial uncertainty about the long-term value of AI progress: While 68.3\% thought good outcomes from superhuman AI are more likely than bad, of these net optimists 48\% gave at least a 5\% chance of extremely bad outcomes such as human extinction, and 59\% of net pessimists gave 5\% or more to extremely \textit{good} outcomes. Between 37.8\% and 51.4\% of respondents gave at least a 10\% chance to advanced AI leading to outcomes as bad as human extinction. More than half suggested that ``substantial'' or ``extreme'' concern is warranted about six different AI-related scenarios, including spread of false information, authoritarian population control, and worsened inequality. There was disagreement about whether faster or slower AI progress would be better for the future of humanity. However, there was broad agreement that research aimed at minimizing potential risks from AI systems ought to be prioritized more. 

\end{abstract}

\newpage

\section{Introduction} \label{sec:introduction}
Artificial intelligence appears poised to reshape society. Decision-makers are working to address opportunities and threats due to AI in the private sector \citep{openai2023moving}, academia \citep{chai2023publications}, and government at the state, national, and international levels \citep{newsom2023executive,  aigov2023,  iawgai2022principles}.

Navigating this situation requires judgments about how the progress and impact of AI are likely to unfold. However, there is a lack of apparent consensus among AI experts on the future of AI \citep{korzekwa2023views}. These judgments are difficult, and there are no established methods of making them well. Thus we must combine various noisy methods, such as extrapolating progress trends \citep{epoch2023scalinglawsliteraturereview}; reasoning about reference classes of similar events \citep{grace2021discontinuous}; analyzing the nature of agents \citep{omohundro2008basic}; probing qualities of current AI systems and techniques \citep{park2023ai}; applying economic models to AI scenarios \citep{jones2023ai, trammell2023economic}; and relying on forecasting aggregation systems such as markets, professional forecasters, and the judgments of various subject matter experts.

One important source of evidence comes from the predictions of AI researchers. Their familiarity with the technology and the dynamics of its past progress puts them in a good position to make educated guesses about the future of AI. However, they are experts in AI research, not AI forecasting and might thus lack generic forecasting skills and experience, or expertise in non-technical factors that influence the trajectory of AI. While AI experts' predictions should not be seen as a reliable guide to objective truth, they can provide one important piece of the puzzle.

We conducted a survey of \numprint{2778} AI researchers who had published peer-reviewed research in the prior year in six top AI venues (NeurIPS, ICML, ICLR, AAAI, IJCAI, JMLR). This to our knowledge constitutes the largest survey of AI researchers to date. The survey took place in the fall of 2023, after an eventful year of broad progress (including the launch of ChatGPT and GPT-4, Google's Bard, Bing AI Chat, Anthropic's Claude 1 and 2), shift in public awareness of AI issues (including two widely signed and publicized AI safety letters \citep{2023openletter, cais2023statement}), and governments beginning to address questions of AI regulation in the US, UK, and EU \citep{biden2023executive, amodei2023testimony, safetysummit, parliament2023eu}. The survey included questions about the speed and dynamics of AI progress, and the social consequences of more advanced AI.

\section{The Survey} \label{sec:survey}
This survey, the ``2023 Expert Survey on Progress in AI,'' or ESPAI, is the third in a series of very similar surveys. The first two were conducted in 2016 \citep{grace2018} and 2022 \citep{grace2022}. The 2023 survey included around four times as many participants as the 2022 survey by expanding from two publication venues (NeurIPS and ICML) to six. It also includes several new questions to probe the nature of future AI systems and diverse potential risks. The survey complements a collection of recent work gathering views on similar questions from the public \citep{steinperlman2023surveys} and corporate leadership \citep{chui2023state}. The full set of questions is available from \cite{survey2023questions}.

Most questions solicited responses in one of three ways: on a Likert scale (multiple choice along a single axis); a probability estimate; or an estimate of a future year. A smaller number of questions asked for write-in responses or numerical estimates.

The way a question is framed can greatly influence the response, and this seems more likely for complex questions \citep{tversky1981framing}. To assess and mitigate framing effects, we often posed different variations of questions on the same topic to different random subsets of respondents. For example, all questions about how soon a milestone would be reached were framed in two ways: fixed-years and fixed-probabilities. Half of respondents were asked to estimate the probability that a milestone would be reached by a given year (``fixed-years framing''), while the other half were asked to estimate the year by which the milestone would be feasible with a given probability (``fixed-probabilities framing''). To minimize confusion, each participant received one framing throughout the survey.

In several parts of the survey, each participant randomly received questions on only one of several topics, to keep the survey brief. We allocated these questions to differently sized subsets of participants based on factors like the importance of the question and the relative value of a larger sample size. This means that most questions were not assigned to all \numprint{2778} participants (see Section~\ref{sec:methods}: Methods).

Several questions asked participants to estimate how many years until a milestone will be feasible. In these questions, we asked participants to provide three year-probability pairs (either via the fixed-years framing or fixed-probabilities framing described above), which we used to approximate a probability distribution for that participant by fitting a gamma cumulative distribution function to these points.

\begin{figure}
\centering
\includegraphics[width=5.5in]{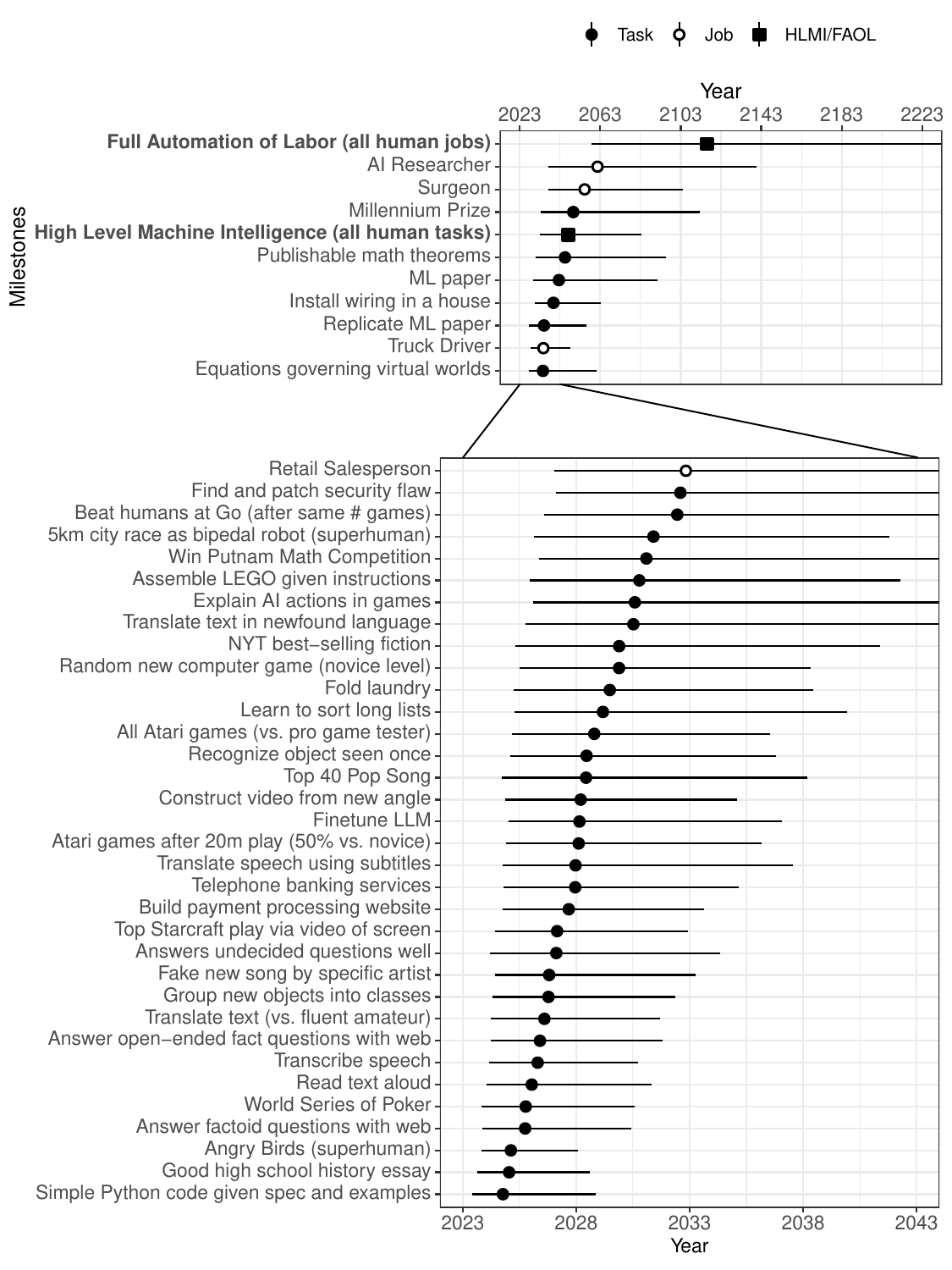}
\caption{\textbf{Most milestones are predicted to have better than even odds of happening within the next ten years, though with a wide range of plausible dates.} The figure shows aggregate distributions over when selected milestones are expected, including 39 tasks, four occupations, and two measures of general human-level performance (see Section~\ref{sec:hlmi}), shown as solid circles, open circles, and solid squares respectively. Circles/squares represent the year where the aggregate distribution gives a milestone a 50\% chance of being met, and intervals represent the range of years between 25\% and 75\% probability. Note that these intervals represent an aggregate of uncertainty expressed by participants, not estimation uncertainty. The displayed milestone descriptions are summaries; for full descriptions, see Appendix~\ref{app:milestone_descriptions}.}
\label{fig:milestones}
\end{figure}

\section{Results on AI Progress}
\subsection{How soon will 39 tasks be feasible for AI?} \label{sec:feasibletasks}
The survey asked about when each of 39 tasks would become feasible, where ``feasible'' was described as meaning ``one of the best resourced labs could implement it in less than a year if they chose to. Ignore the question of whether they would choose to.'' Each respondent was asked about four tasks, so that each task received around 250 estimates. Each respondent gave three probability-year pairs per task.

To aggregate the responses, we first fit a gamma distribution to each participant's three probability-year pairs, and then computed the mean across the participants' individual gamma distributions.

All but four of the 39 tasks were predicted to have at least a 50\% chance of being feasible within the next ten years (Figure~\ref{fig:milestones}). This includes several economically very valuable tasks---such as coding an entire payment processing site from scratch and writing new songs indistinguishable from real ones by hit artists such as Taylor Swift. It also includes tasks that imply substantial progress in sample-efficiency (e.g. ‘Beat novices in 50\% of Atari games after 20 minutes of play’), AI-driven AI progress (e.g. autonomously fine-tuning an open-source LLM), and robotics (e.g. folding laundry). 

The six tasks expected to take longer than ten years were: ``After spending time in a virtual world, output the differential equations governing that world in symbolic form'' (12 years), ``Physically install the electrical wiring in a new home'' (17 years), ``Research and write'' (19 years) or ``Replicate'' (12 years) ``a high-quality ML paper,'' ``Prove mathematical theorems that are publishable in top mathematics journals today'' (22 years), and solving ``long-standing unsolved problems in mathematics'' such as a Millennium Prize problem (27 years).

\subsubsection{Comparison with 2022}

\begin{figure}
\centering
\includegraphics[width=5in]{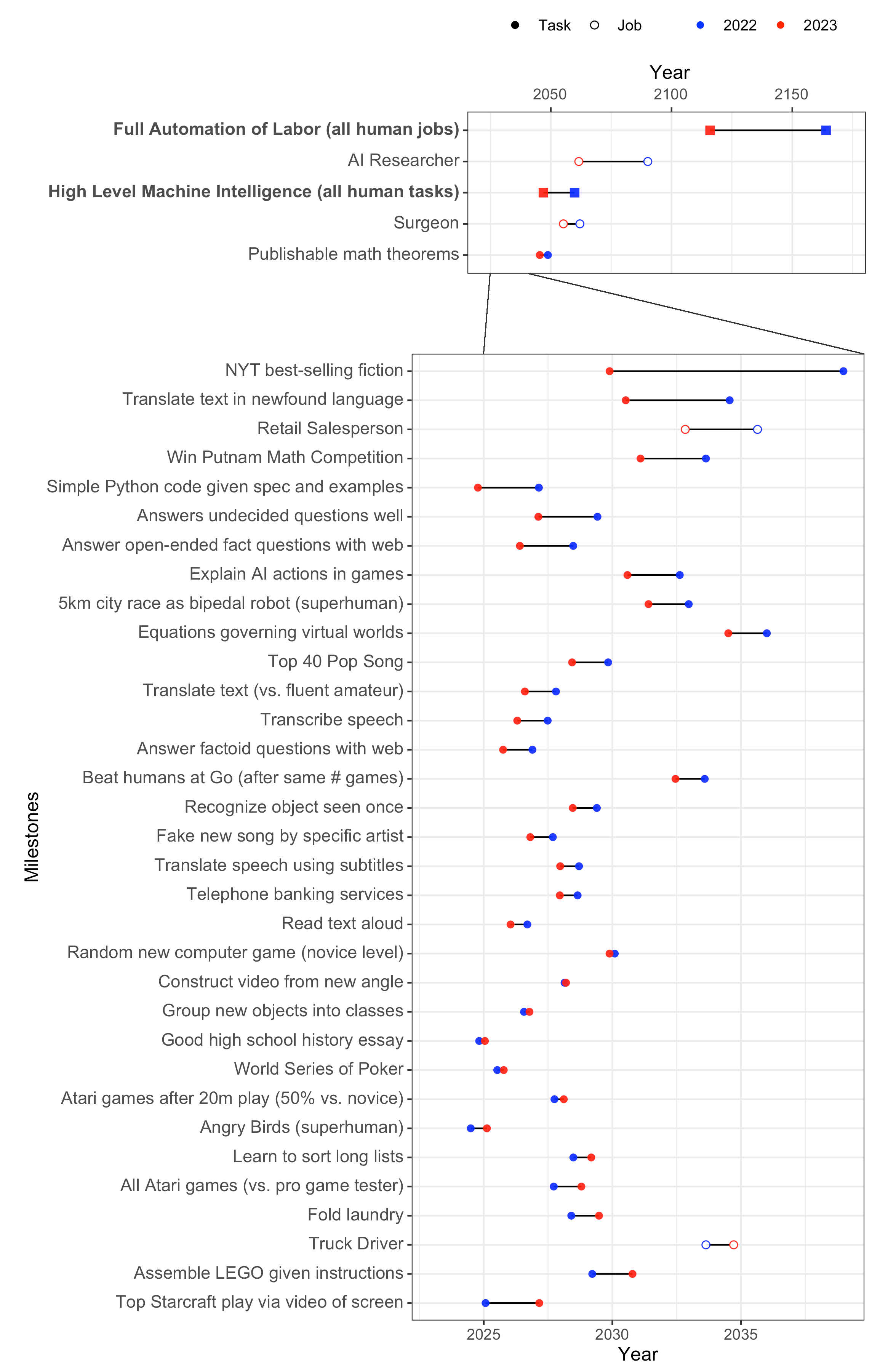}
\caption{\textbf{Expected feasibility of many AI milestones moved substantially earlier in the course of one year (between 2022 and 2023).} The milestones are sorted (within each scale-adjusted chart) by size of drop from 2022 forecast to 2023 forecast, with the largest change first. The year when the aggregate distribution gives a milestone a 50\% chance of being met is represented by solid circles, open circles, and solid squares for tasks, occupations, and general human-level performance respectively. The three groups of questions have different formats that may also influence answers. For full descriptions of the summarized milestones, see Appendix~\ref{app:milestone_descriptions}.}
\label{fig:milestoneshift}
\end{figure}

32 AI task questions were identical to those in the 2022 survey, as well as the 2016 survey \citep{grace2018}, with the exception of minor edits to task descriptions for updated accuracy between 2016 and 2022 \citep{survey2023wikipage}. All tasks from \cite{grace2018} were included, regardless of whether the authors would judge them to be achieved. The survey population included more conferences in 2023, but this did not appear to have a notable effect on opinion (See~\ref{sec:samplechange}). Figure~\ref{fig:milestoneshift} shows how expected dates for reaching these milestones shifted from 2022 to 2023.

Between 2022 and 2023, aggregate predictions for 21 out of 32 tasks moved earlier. The aggregate predictions for 11 tasks moved later.

On average, for the 32 tasks included in both the 2022 and 2023 surveys, the 50\textsuperscript{th} percentile year they were expected to become feasible shifted 1.0 years earlier (SD = 2.0, SE = 0.18).

\subsection{How soon will human-level performance on all tasks or occupations be feasible?} \label{sec:hlmi}
We asked how soon participants expected AI systems to outperform humans across all activities, which we framed in two ways: as either \textit{tasks}, in the question about ``High-Level Machine Intelligence'' (HLMI), or \textit{occupations}, in the question about ``Full Automation of Labor'' (FAOL).

\subsubsection{How soon will `High-Level Machine Intelligence' be feasible?}
We defined High-Level Machine Intelligence (HLMI) thus:

\begin{quote}
High-level machine intelligence (HLMI) is achieved when unaided machines can accomplish every task better and more cheaply than human workers. Ignore aspects of tasks for which being a human is intrinsically advantageous, e.g. being accepted as a jury member. \textit{Think feasibility, not adoption}.
\end{quote}

We asked for predictions, assuming ``human scientific activity continues without major negative disruption.'' We aggregated the results (n=\numprint{1714}) by fitting gamma distributions, as with individual task predictions in~\ref{sec:feasibletasks}.

\begin{figure}
\centering
\includegraphics[width=4.5in]{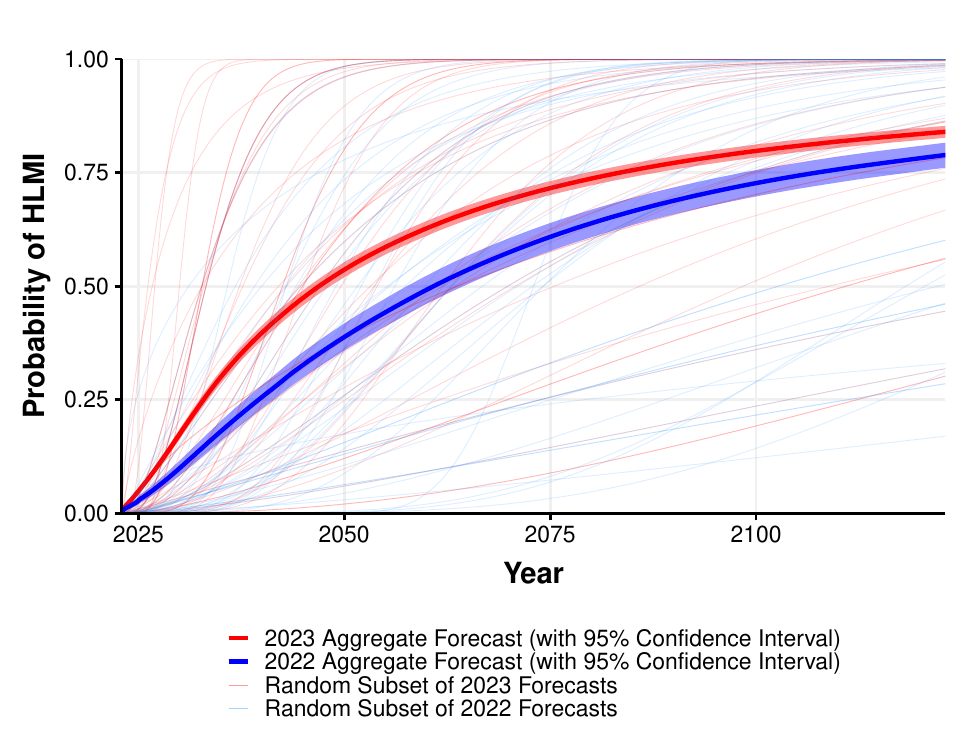}
\includegraphics[width=4.5in]{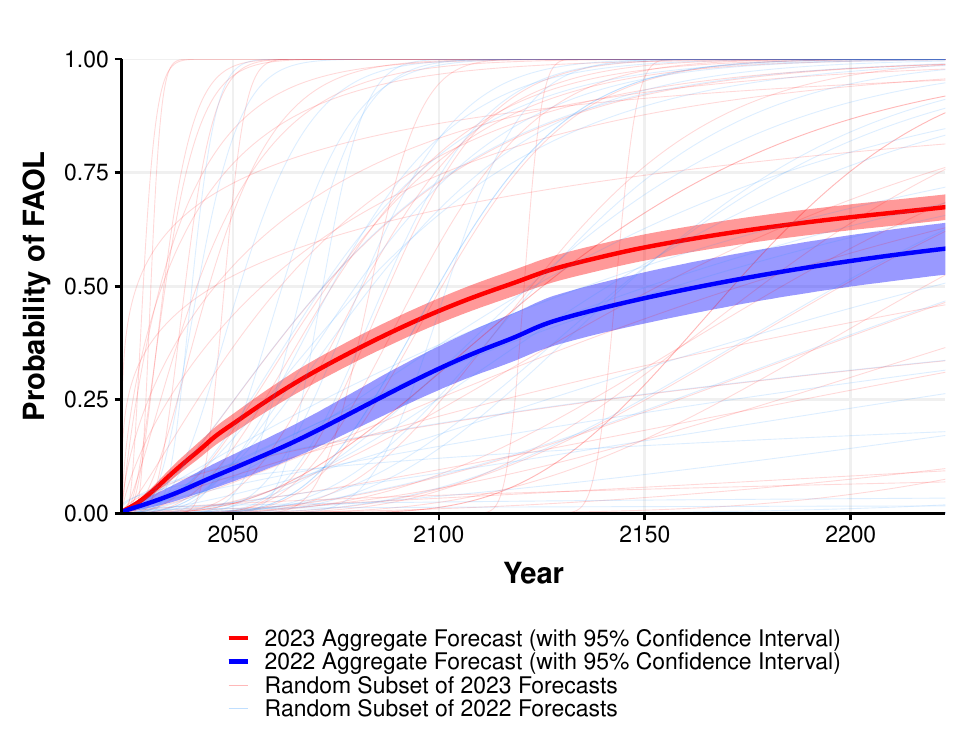}
\caption{\textbf{Aggregate forecast for 50\textsuperscript{th} percentile arrival time of High-Level Machine intelligence (HLMI) dropped by 13 years between 2022 and 2023. The forecast for 50\textsuperscript{th} percentile arrival time of Full Automation of Labor (FAOL) dropped by 48 years in the same period. However, there is still a lot of uncertainty.} ``Aggregate Forecast'' is the mean distribution over all individual cumulative distribution functions. For comparison, we included the 2022 Aggregate Forecast. To give a sense of the range of responses, we included random subsets of individual 2023 and 2022 forecasts. Note that the thinner `confidence interval` in 2023 (compared to 2022) is due to our increased confidence about the average respondents' views due to a larger sample size, not respondents' predictions converging.}
\label{fig:hlmiarrival}
\label{fig:faol}
\end{figure}

In both 2022 and 2023, respondents gave a wide range of predictions for how soon HLMI will be feasible (Figure~\ref{fig:hlmiarrival}). The aggregate 2023 forecast predicted a 50\% chance of HLMI by 2047, down thirteen years from 2060 in the 2022 survey. For comparison, in the six years between the 2016 and 2022 surveys, the expected date moved only one year earlier, from 2061 to 2060\footnote{Reported as 2059 in 2022; small code changes and improvements to data cleaning between surveys shifted the aggregate slightly.}.

The aggregate 2023 forecast predicted a 10\% chance of HLMI by 2027, down two years from 2029 in the 2022 survey. See Appendix~\ref{app:supplementary_figures} for a table comparing this with Full Automation of Labor, which is discussed below.

\subsubsection{How soon will `Full Automation of Labor' be feasible?}
The other framing of the question about how soon AI systems would outperform humans across all activities was about ``Full Automation of Labor,'' or FAOL. We defined FAOL thus:

\begin{quote}
Say an occupation becomes fully automatable when unaided machines can accomplish it better and more cheaply than human workers. Ignore aspects of occupations for which being a human is intrinsically advantageous, e.g. being accepted as a jury member. \textit{Think feasibility, not adoption.} [\ldots]

Say we have reached `full automation of labor' when all occupations are fully automatable. That is, when for any occupation, machines could be built to carry out the task better and more cheaply than human workers.
\end{quote}

Before the participants (n=\numprint{774}) were asked about the full automation of labor, respondents were asked when four specific occupations would become fully automatable: ``Truck driver,'' ``Surgeon,'' ``Retail salesperson,'' and ``AI researcher'' (Figure~\ref{fig:milestones}). They were also asked to think of an existing human occupation that they thought would be among the final ones to be fully automatable. They were then asked when `full automation of labor' (FAOL) would be achieved.

The aggregate 2023 forecast predicted a 50\% chance of FAOL by 2116, down 48 years from 2164 in the 2022 survey (Figure~\ref{fig:faol}). We checked if this difference was significant for participants who received the question in the fixed-probabilities framing, and found that it was (p = .0052, Yuen's test (bootstrap version); see Appendix~\ref{app:yuen}). There is about a 70-year difference between the mean 50\% prediction for HLMI and the mean 50\% prediction for FAOL. We discuss this surprising finding in the next section, ``Framing effect of HLMI vs FAOL.'' 

Compared to 2016, 2023 has earlier 50\% estimates but later 10\% estimates. (See Appendix~\ref{app:supplementary_figures})

The answers to the write-in question about an existing occupation likely to be among the last automatable were categorized according to O*NET's \textit{All Job Family Occupations} categories \citep{onetcategories}.  The top five most-suggested categories were: ``Computer and Mathematical'' (91 write-in answers in this category), ``Life, Physical, and Social Science'' (77 answers), ``Healthcare Practitioners and Technical'' (56), ``Management'' (49), and ``Arts, Design, Entertainment, Sports, and Media'' (39).

The answers to the write-in question about an existing occupation likely to be among the last automatable were categorized according to O*NET’s categories \citep{onetcategories}. The top five most-suggested categories were: ``Computer and Mathematical'' (91 write-in answers in this category), ``Life, Physical, and Social Science'' (77 answers), ``Healthcare Practitioners and Technical'' (56), ``Management'' (49), and ``Arts, Design, Entertainment, Sports, and Media'' (39).

\subsubsection{Differences between HLMI and FAOL}

Predictions for a 50\% chance of the arrival of FAOL are consistently more than sixty years later than those for a 50\% chance of the arrival of HLMI. This was seen in the results from the surveys of 2023, 2022, and 2016. This is surprising because HLMI and FAOL are quite similar: FAOL asks about the automation of all occupations; HLMI asks about the feasible automation of all tasks. Since occupations might naturally be understood either as complex tasks, composed of tasks, or closely connected with one of these, achieving HLMI seems to either imply having already achieved FAOL, or suggest being close.

We do not know what accounts for this gap in forecasts. Insofar as HLMI and FAOL refer to the same event, the difference in predictions about the time of their arrival would seem to be a framing effect.

However, the relationship between ``tasks'' and ``occupations'' is debatable. And the question sets do differ beyond definitions: only the HLMI questions are preceded by the instruction to ``assume that human scientific activity continues without major negative disruption,'' and the FAOL block asks a sequence of questions about the automation of specific occupations before asking about full automation of labor. So conceivably this wide difference could be caused by respondents expecting major disruption to scientific progress, or by the act of thinking through specific examples shifting overall anticipations. From our experience with question testing, it also seems possible that the difference is due to other differences in interpretation of the questions, such as thinking of automating occupations but not tasks as including physical manipulation, or interpreting FAOL to require \textit{adoption} of AI in automating occupations, not mere \textit{feasibility} (contrary to the question wording).

\subsubsection{Demographic differences}
Geographical background was correlated with expectations about the timing of human-level performance: respondents whose undergraduate education was in Asia anticipated an 11 year earlier arrival of HLMI than participants from Europe, North America, or other regions combined. See Appendix~\ref{app:demographics} for more demographic comparisons.

\subsubsection{Do participants think they agree on timing of HLMI?}
We asked respondents a set of ``meta'' questions about their views on others' views (n=\numprint{671}). One meta question asked to what extent they thought that they disagreed with the typical AI researcher about when HLMI would exist. 44\% said ``Not much,'' 46\% said ``A moderate amount,'' and 10\% said ``A lot.''

\subsection{Framing effect of fixed-years vs fixed-probabilities}  \label{sec:framings}
All questions about how soon a milestone would be reached were framed in two ways: fixed-years and fixed-probabilities. In either framing, we ask for three year-probability pairs, but in one we fix a set of probabilities (10\%, 50\%, 90\%) and ask how many years until the participant would assign each probability to the milestone being met, whereas in the other framing we fix a set of future years (usually 10 years, 20 years, 50 years) and ask about the probability of the milestone occurring by that year.

The fixed-years framing has been previously observed to produce systematically later predictions \citep{grace2018,grace2022}, but we do not know if one framing is more accurate than the other. Here we have used both and combined them with equal weight.

The previously-observed framing effect was again observed in this survey. For example, the year with a 50\% chance of HLMI from participants answering in the fixed-year frame (34 years) was twice as far into the future as that for participants answering in the fixed-probability frame (17 years). However, it's notable that even the larger of these two is shorter than 2022’s combined forecast (37 years), demonstrating a substantial shift of predictions closer to the present (Fig~\ref{fig:hlmi_fixed_years_framing} in Appendix~\ref{app:supplementary_figures}).

\begin{figure}[h]
\centering
\includegraphics[width=4.5in]{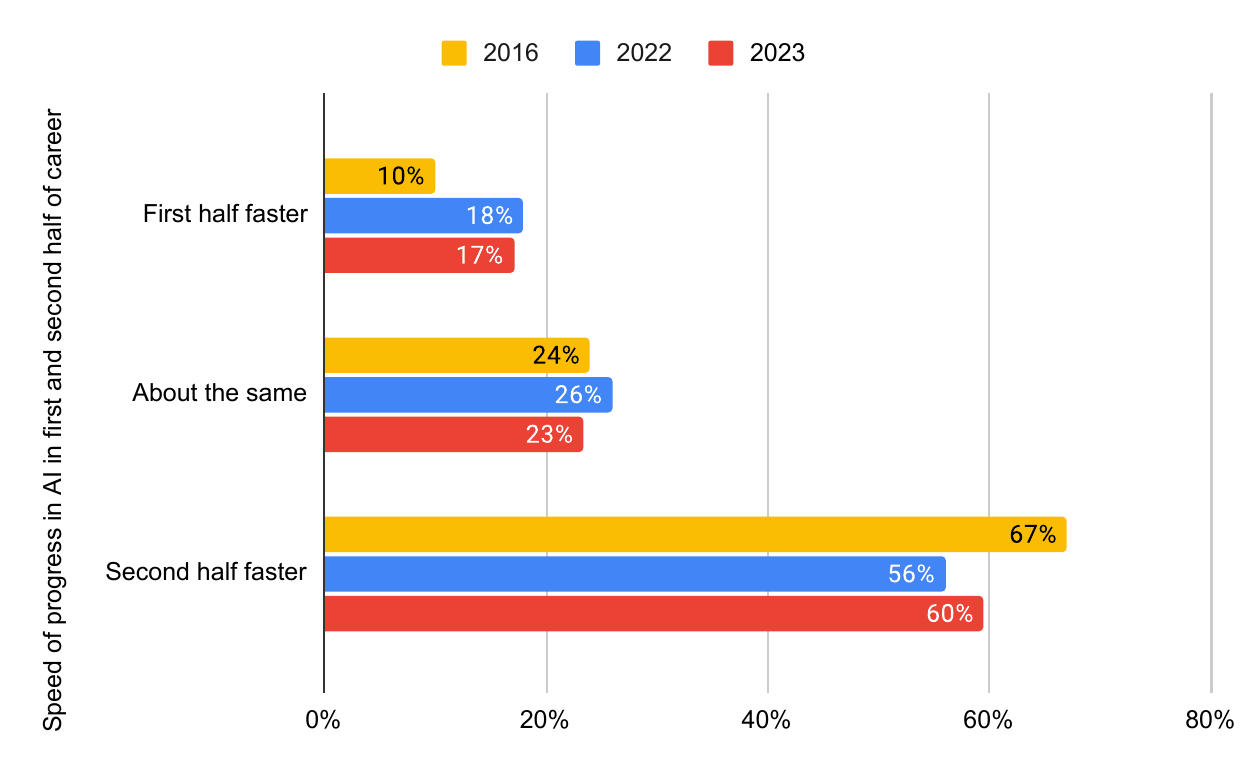}
\caption{\textbf{Most respondents indicated that the pace of progress in their area of AI increased between the first and second half of their time in a field.} Participants were asked whether the second half of the time they had spent working in their area of AI saw more progress than the first half. The median time working in the area was 5 years.}
\label{fig:careerspeed}
\end{figure}

\subsection{Change in observed rates of progress}
We asked respondents which AI area they had worked in for the longest and whether progress in the second half was faster than the first (Figure~\ref{fig:careerspeed}).

\subsection{What causes AI progress?}
\begin{figure}[h]
\centering
\includegraphics[width=3.5in]{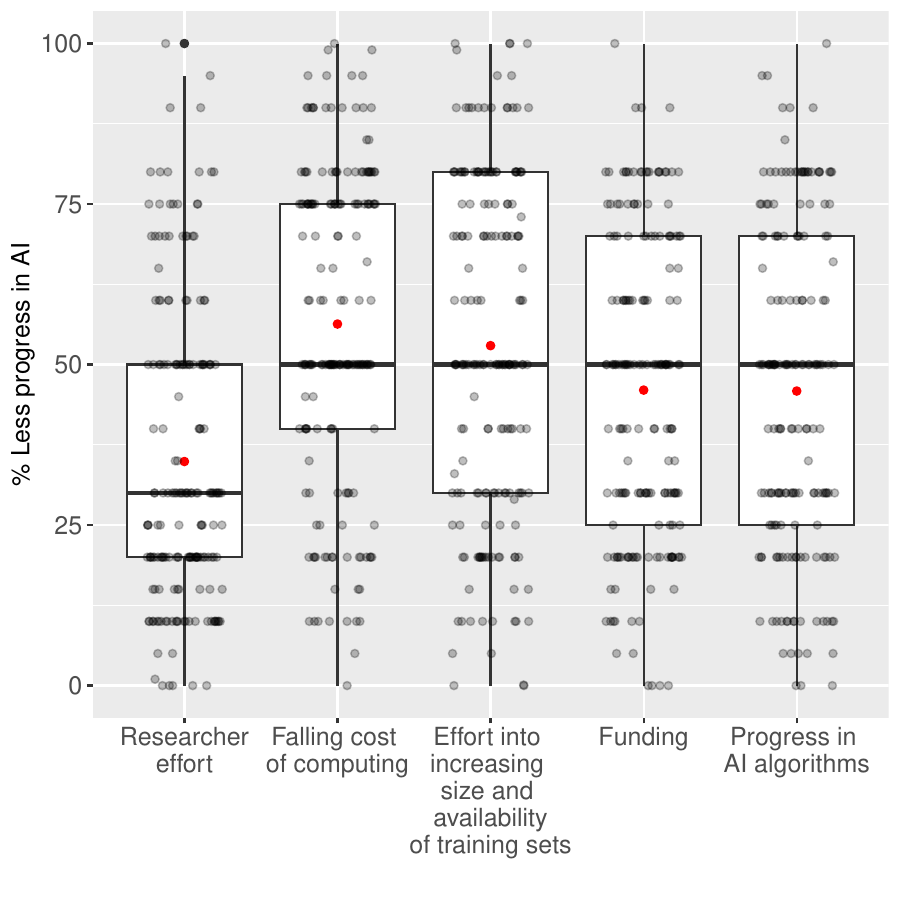}
\caption{\textbf{Estimated reduction in AI progress if inputs had been halved over the past decade.} Red dots represent means. Boxes contain the 25\textsuperscript{th} to 75\textsuperscript{th} percentile range; middle lines are medians. Whiskers are the least and greatest values that are not more than 1.5 times the interquartile range from the median. Participants estimated that halving the drop in costs of computing would have had the greatest effect on AI progress over the last decade, while halving `researcher effort' and `progress in AI algorithms' would have had the least effect. Overall, all the included inputs were seen as having contributed substantially to AI progress.}
\label{fig:sensitivity_of_progress}
\end{figure}

We asked about the sensitivity of progress in AI capabilities to changes in five inputs: 1) researcher effort, 2) decline in cost of computation, 3) effort put into increasing the size and availability of training datasets, 4) funding, and 5) progress in AI algorithms. We asked respondents to imagine that only half as much of each input had been available over the past decade,\footnote{The declining cost of computation was an exception; here it was framed as ``over the last \textit{n} years,'' and the question was about costs falling around half as far on a log scale. This was done to match previous surveys.} and the effect they would expect this to have had on the rate of AI progress. The results are shown in Figure~\ref{fig:sensitivity_of_progress}.

There was a wide range of views about each input, implying a large degree of uncertainty. Additionally, the relatively even distribution of predictions cuts against a common narrative that progress in cheap computing is the dominant driver of AI progress. Across all inputs, we observe many more answers of ``0\%'' (no difference) and ``100\%'' (all AI progress lost) than we would expect, which suggests to us possible misunderstandings of the question.

\subsection{Will there be an intelligence explosion?}
We asked respondents about the possibility, after HLMI is hypothetically achieved, of an `intelligence explosion,' as explained in this question:

\begin{quote}
Some people have argued the following:
\begin{quote}
If AI systems do nearly all research and development, improvements in AI will accelerate the pace of technological progress, including further progress in AI.

Over a short period (less than 5 years), this feedback loop could cause technological progress to become more than an order of magnitude faster.
\end{quote}

How likely do you find this argument to be broadly correct?
\end{quote}

\begin{figure}
\centering
\includegraphics[width=4.5in]{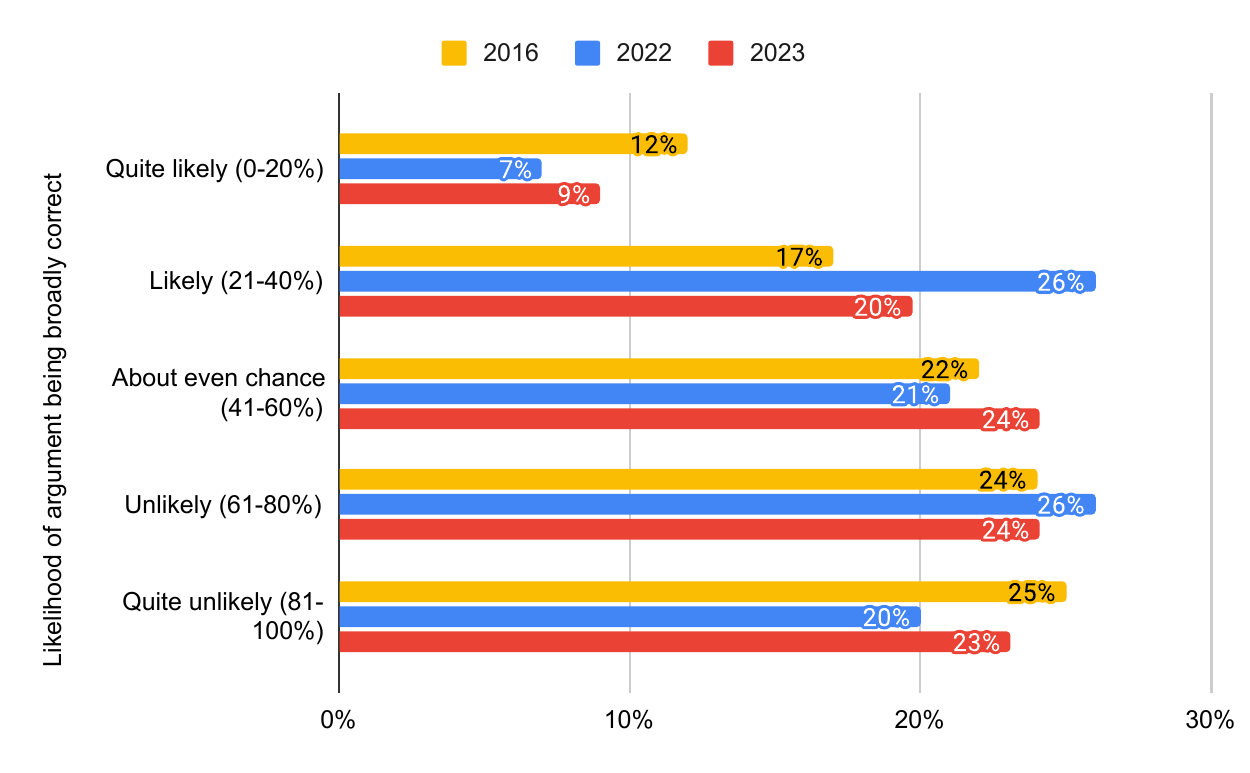}
\caption{\textbf{Since 2016 a majority of respondents have thought that it's either ``quite likely,'' ``likely,'' or an ``about even chance'' that technological progress becomes more than an order of magnitude faster within 5 years of HLMI being achieved.}}
\label{fig:explosion_correct}
\end{figure}

The results to this first question are shown in Figure~\ref{fig:explosion_correct}.

The other two `intelligence explosion' questions were about the likelihood of a dramatically increased rate of global technological advancement 2 and 30 years post-HLMI, and AI that can outperform humans across all professions 2 and 20 years post-HLMI. Results are shown in Table~\ref{tab:explosive_progress}.
\renewcommand{\arraystretch}{1.5}

\begin{table}
    \centering
    \begin{tabular}{r|ccc}
         & \multicolumn{3}{c}{Median probability (N)} \\
         & 2016 & 2022 & 2023 \\
    \hline
    Explosive progress 2 years after HLMI & 20\% (225) & 20\% (339) & 20\% (298) \\
    Explosive progress 30 years after HLMI & 80\% (225)  & 80\% (339) & 80\% (297) \\
    Intelligence explosion argument is broadly correct & ``41-60\%'' (232) & ``41-60\%'' (386) & ``41-60\%'' (299) \\
    AI is vastly better than humans 2 years after HLMI & 10\% (213) & 10\% (371) & 10\% (281) \\
    AI is vastly better than humans 30 years after HLMI & 50\% (214) & 60\% (371) & 60\% (282)\\
    \hline
    \end{tabular}
    \caption{\textbf{Results of three questions regarding a hypothetical intelligence explosion have remained remarkably stable since 2016.}}
    \label{tab:explosive_progress}
\end{table}
In sum, across these three questions, the median participant did not overall expect something like a rapid acceleration of progress from an `intelligence explosion,' but did give substantial credence to it.

Further figures related to intelligence explosion questions are in Appendix~\ref{app:supplementary_figures}.

\subsection{What will AI systems in 2043 be like?} \label{sec:ai2043}
Concerns about risks from future AI systems are often linked to specific traits related to alignment, trustworthiness, predictability, self-directedness, capabilities, and jailbreakability. We asked respondents how likely it was that at least some state-of-the-art AI systems in 2043 would have each of eleven such traits ($n \in \left[649,667\right]$\footnote{i.e. the number of people (n) who answered each of these eleven questions was between 649 and 667}).

\begin{figure}
\centering
\includegraphics[scale=0.3]{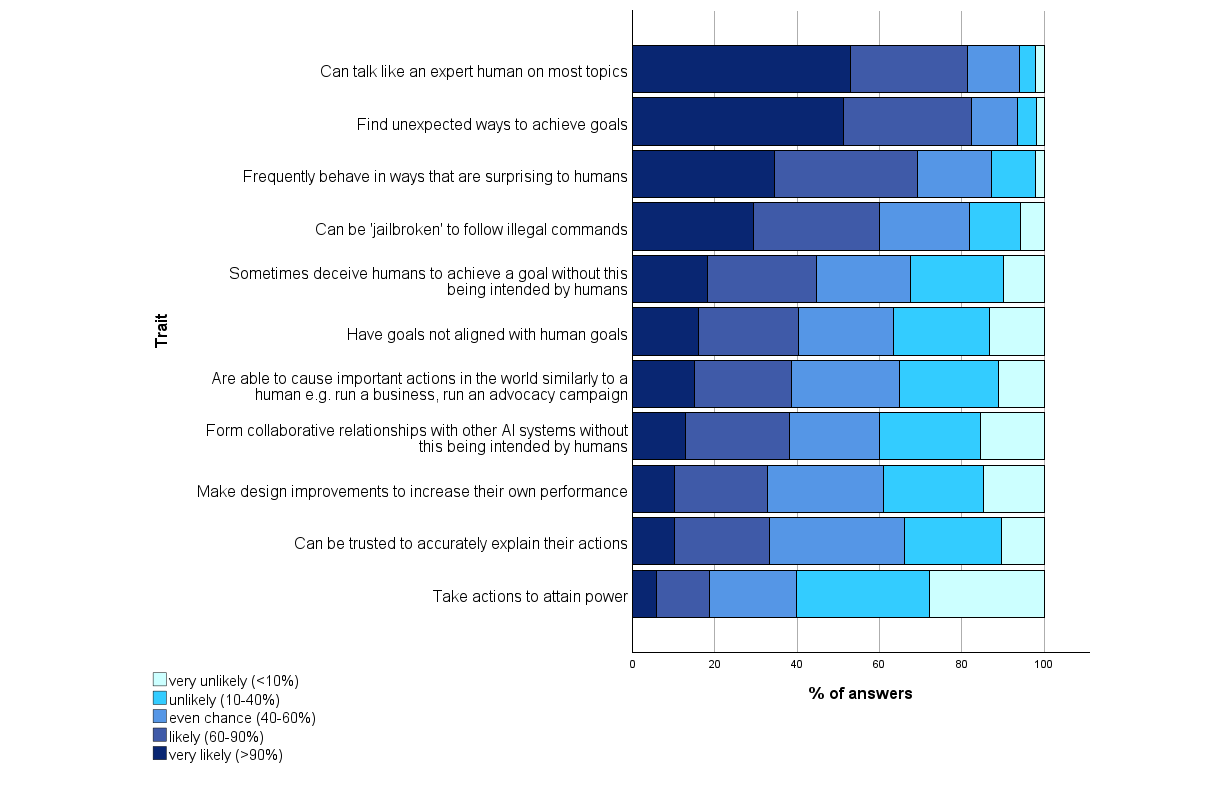}
\caption{\textbf{Respondents' estimates of the likelihood that at least some AI systems in 2043 will have each of these traits; organized from least to most likely.}}
\label{fig:traits2043}
\end{figure}

All 11 traits were considered to have a relatively high chance of existing in AI systems in 2043, though with much uncertainty. Only one trait had a median answer below `even chance': ``Take actions to attain power.'' While there was no consensus even on this trait, it's notable that it was deemed least likely, because it is arguably the most sinister, being key to an argument for extinction-level danger from AI \citep{carlsmith2022power}.

Answers reflected substantial uncertainty and disagreement among participants. No trait attracted near-unanimity on any probability, and no more than 55\% of respondents answered ``very likely'' or ``very unlikely'' about any trait. (Figure~\ref{fig:traits2043})

There were areas of agreement, however. For instance, a large majority of participants thought state-of-the-art AI systems in twenty years would be likely or very likely to:

\begin{enumerate}
\item Find unexpected ways to achieve goals (82.3\% of respondents),
\item Be able to talk like a human expert on most topics (81.4\% of respondents), and
\item Frequently behave in ways that are surprising to humans (69.1\% of respondents)
\end{enumerate}

\subsection{Will AI in 2028 truthfully and intelligibly explain its decisions?}

\begin{figure}[h]
\centering
\includegraphics[width=4in]{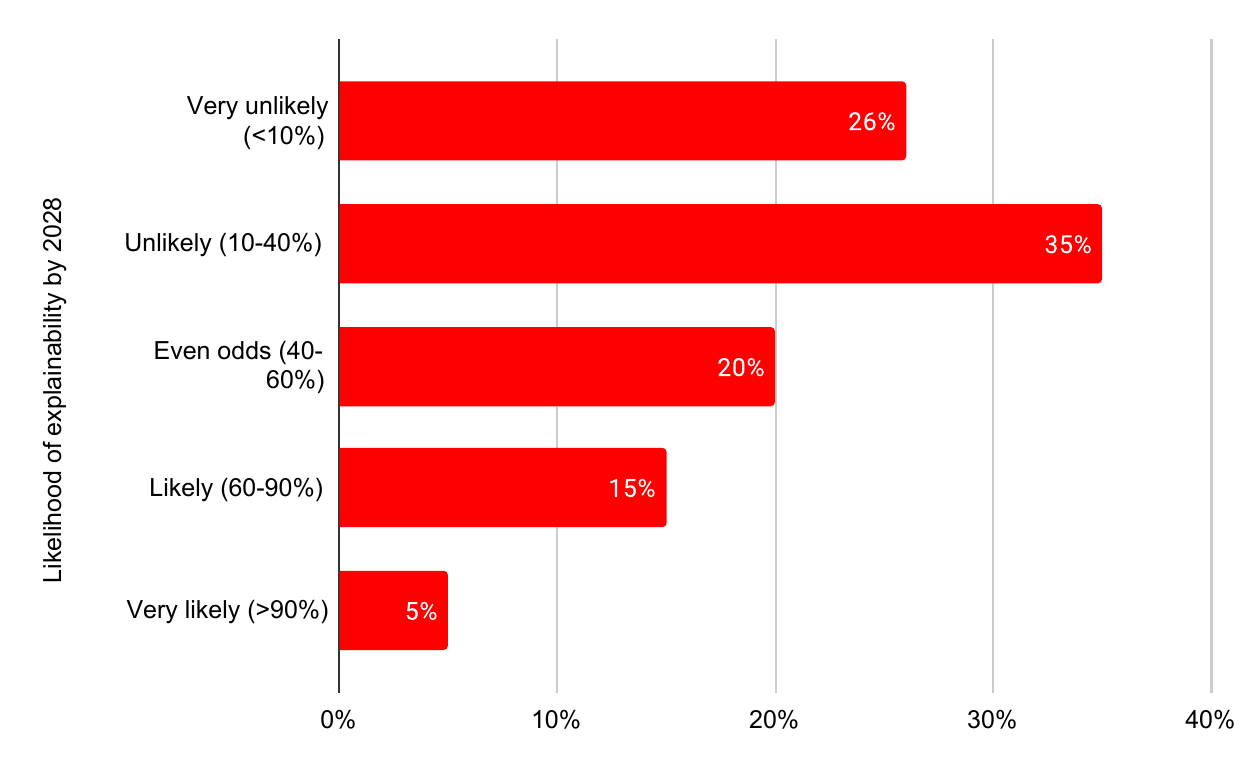}
\caption{\textbf{Most respondents considered it unlikely that users of AI systems in 2028 will be able to know the true reasons for the AI systems' choices, with only 20\% giving it better than even odds. (n=912)}}
\label{fig:explainability}
\end{figure}

Uninterpretable reasoning in AI systems is often considered an AI risk factor, potentially leading to outcomes ranging from unjust biases in treatment of people to active pursuit of harm hidden by capable agents. We thus asked about the interpretability of AI systems in five years (Figure~\ref{fig:explainability}):

\begin{quote}
For typical state-of-the-art AI systems in 2028, do you think it will be possible for users to know the true reasons for systems making a particular choice? By ``true reasons'' we mean the AI correctly explains its internal decision-making process in a way humans can understand. By ``true reasons'' we do \textbf{not} mean the decision itself is correct.
\end{quote}

This is related to the question in Section~\ref{sec:ai2043}, which asked how likely it was that at least some state-of-the-art AI systems in 2043 (fifteen years later), ``can be trusted to accurately explain their actions.'' The median answer there was ``even chance'' (40-60\% likely), compared to ``unlikely'' (10-40\%) on this question.
\FloatBarrier

\section{Results on Social Impacts of AI}

\subsection{How concerning are 11 future AI-related scenarios?}
\begin{figure}
\centering
\includegraphics[scale=0.3]{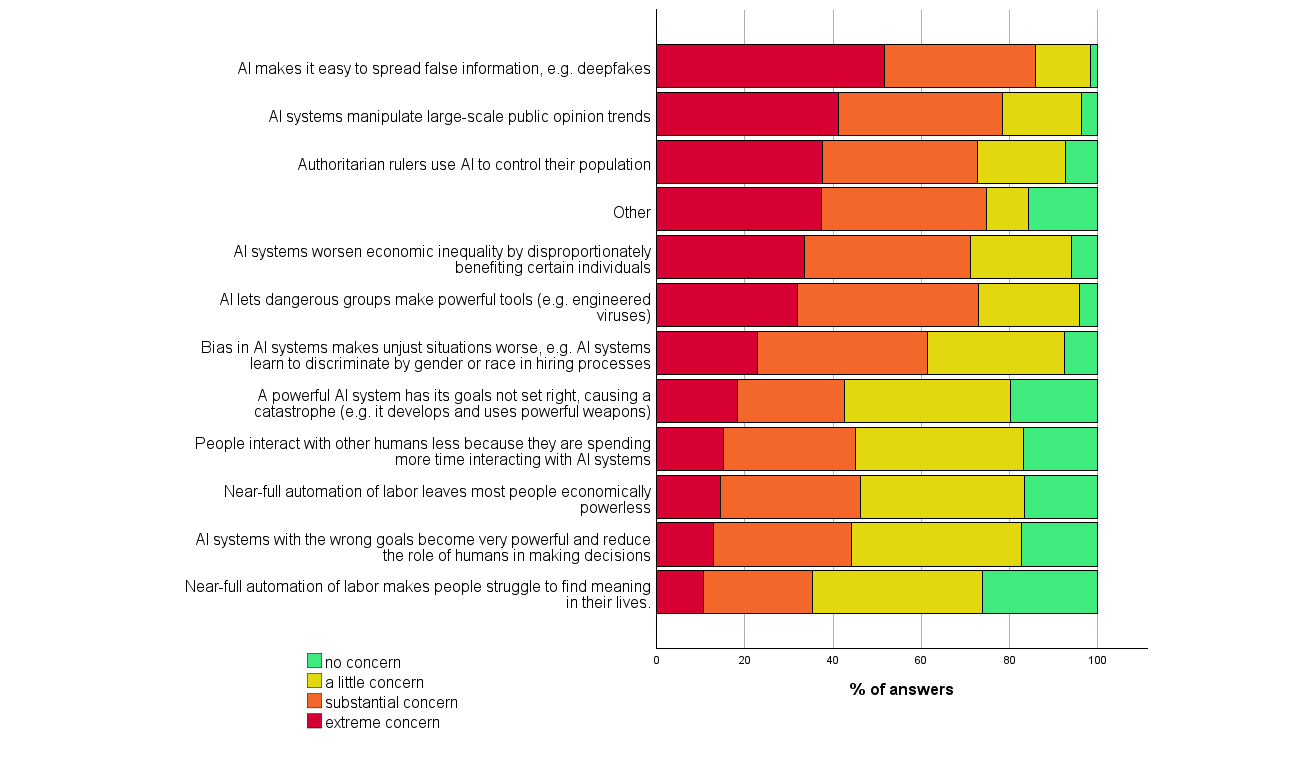}
\caption{\textbf{Amount of concern potential scenarios deserve, organized from most to least extreme concern.}}
\label{fig:negativeoutcomes}
\end{figure}

We asked participants (n=\numprint{1345}) about eleven potentially concerning AI scenarios, such as AI-enabled misinformation, worsened economic inequality, and biased AI systems worsening injustice. We asked how much concern each deserved in the next thirty years (Figure~\ref{fig:negativeoutcomes}).

Each scenario was considered worthy of either substantial or extreme concern by more than 30\% of respondents. As measured by the percentage of respondents who thought a scenario constituted either a ``substantial'' or ``extreme'' concern, the scenarios worthy of most concern were: spread of false information e.g. deepfakes (86\%), manipulation of large-scale public opinion trends (79\%), AI letting dangerous groups make powerful tools (e.g. engineered viruses) (73\%), authoritarian rulers using AI to control their populations (73\%), and AI systems worsening economic inequality by disproportionately benefiting certain individuals (71\%).

There is some ambiguity about the reason why a scenario might be considered concerning: it might be considered especially disastrous, or especially likely, or both. From our results, there's no way to disambiguate these considerations.

\subsection{How good or bad for humans will High-Level Machine Intelligence be?} \label{sec:goodness_of_hlmi}
We asked participants to assume that, at some point, ``high-level machine intelligence'' (HLMI) will exist, as defined in Section~\ref{sec:hlmi}. Given this assumption for the sake of the question, we asked how good or bad they expect the overall impact of this to be ``in the long run'' for humanity.

\begin{figure}
\centering
\includegraphics[width=6in]{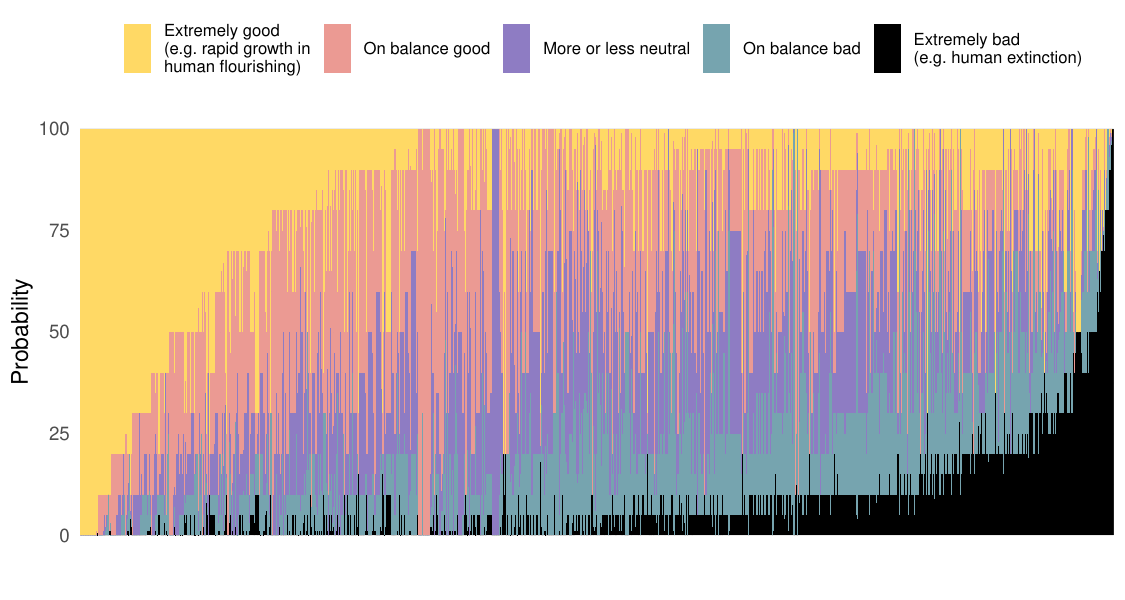}
\caption{\textbf{Respondents exhibited diverse views on the expected goodness/badness of High Level Machine Intelligence (HLMI).} We asked participants to assume, for the sake of the question, that HLMI will be built at some point. The figure shows a random selection of 800 responses on the positivity or negativity of long-run impacts of HLMI on humanity. Each vertical bar represents one participant and the bars are sorted left to right by a weighted sum of probabilities corresponding to overall optimism. Responses range from extremely optimistic to extremely pessimistic. Over a third of participants (38\%) put at least a 10\% chance on extremely bad outcomes (e.g. human extinction).}
\label{fig:ai_goodness}
\end{figure}

Respondents exhibited diverse views on the future impact of advanced AI (Figure~\ref{fig:ai_goodness}), highlighting how respondents' views are more complex than can be represented by an `optimism vs pessimism` axis. Many people who have high probabilities of bad outcomes also have high probabilities of good outcomes. A majority spread their credence across the entire spectrum of outcomes, with 64\% assigning non-zero probabilities to both extremely good and extremely bad scenarios. 68.3\% of participants found good outcomes more likely than bad outcomes, while 57.8\% considered extremely bad outcomes (e.g. human extinction) a nontrivial possibility ($\geq 5\%$ likely). Even among net optimists, nearly half (48.4\%) gave at least 5\% credence to extremely bad outcomes, and among net pessimists, more than half (58.6\%) gave at least 5\% to extremely good outcomes. The broad variance in credence in catastrophic scenarios shows there isn't strong evidence understood by all experts that this kind of outcome is certain or implausible.

The median prediction for extremely bad outcomes, such as human extinction, was 5\% (mean 9\%). Over a third of participants (38\%) put at least a 10\% chance on extremely bad outcomes. This is comparable to, but somewhat lower than, rates of assigning at least 10\% to extinction-level outcomes in answers to other questions more directly about extinction, between 41\% and 51\% (see Section~\ref{sec:extinction}). On the very pessimistic end, one in ten participants put at least a 25\% chance on outcomes in the range of human extinction).

\begin{figure}
\centering
\includegraphics[width=3.2in]{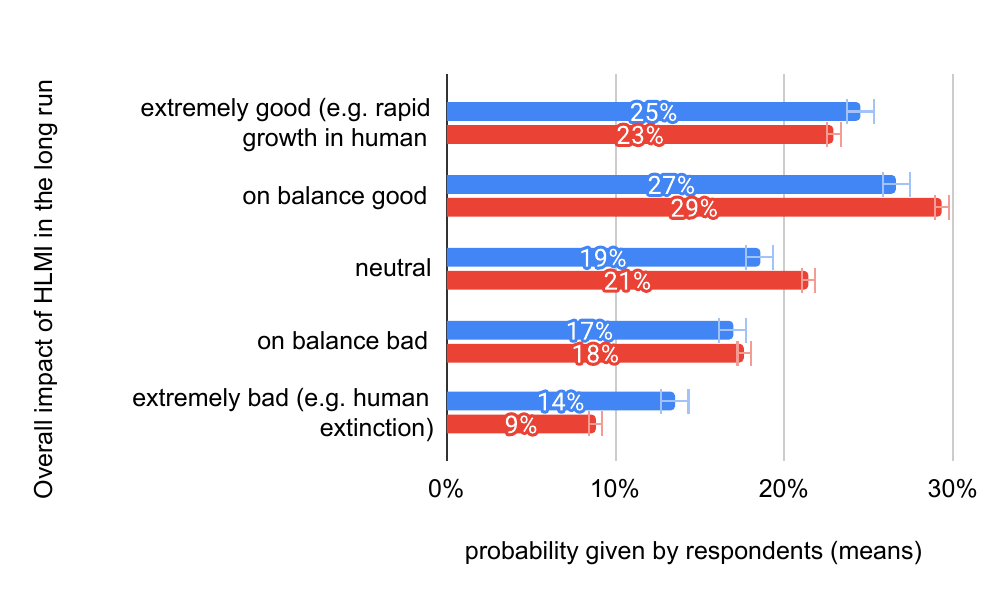}
\includegraphics[width=3.2in]{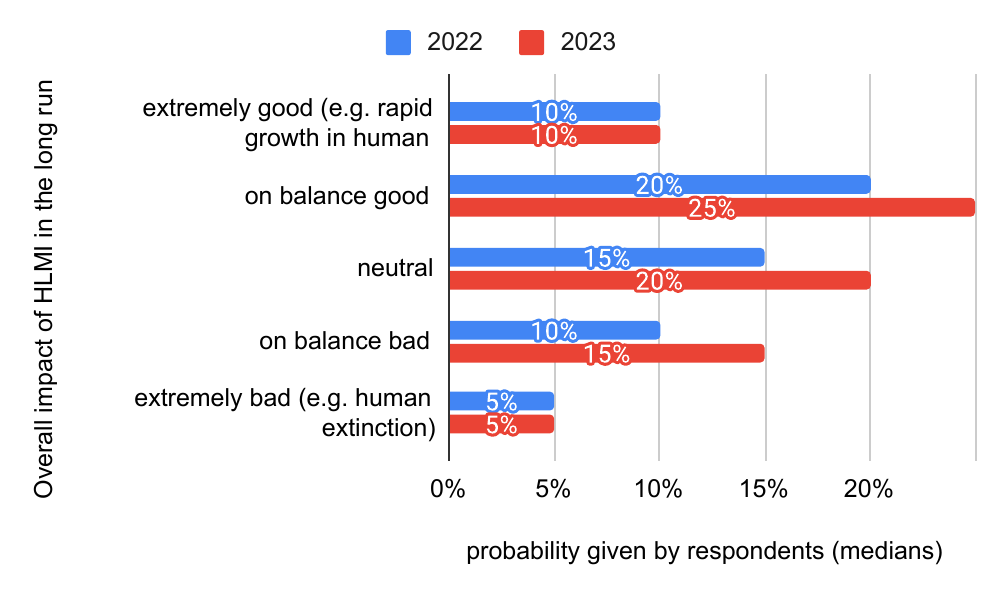}
\caption{\textbf{Mean but not median predictions in 2023 (n=2704) about the consequences of HLMI have shifted slightly away from extreme outcomes compared to 2022 (n=559). Error bars indicate the standard error.}}
\label{fig:hlmi_value}
\end{figure}

Since 2022, mean overall probability on extreme outcomes (good \textit{or} bad) has fallen slightly (Figure~\ref{fig:hlmi_value}). The proportion of people who put at least a 10\% chance on extremely bad outcomes (e.g. human extinction) has fallen from 48\% in 2022 in 2023, and the mean prediction for this type of outcome is down from 14\% to 9.0\%.

Appendix~\ref{app:demographics} contains comparisons between results from different demographics on this question, and Appendix~\ref{app:supplementary_figures} contains another relevant figure.

\subsection{How likely is AI to cause human extinction?} \label{sec:extinction}

To further clarify views on the ``extremely bad (e.g. human extinction)'' scenarios in the question on overall impacts, participants were given one of three similar questions about human extinction. Their differences were intended to help isolate exactly how concerning different scenarios are, what respondents expect to happen, and how much difference working makes.

\begin{figure}
\centering
\includegraphics[width=6in]{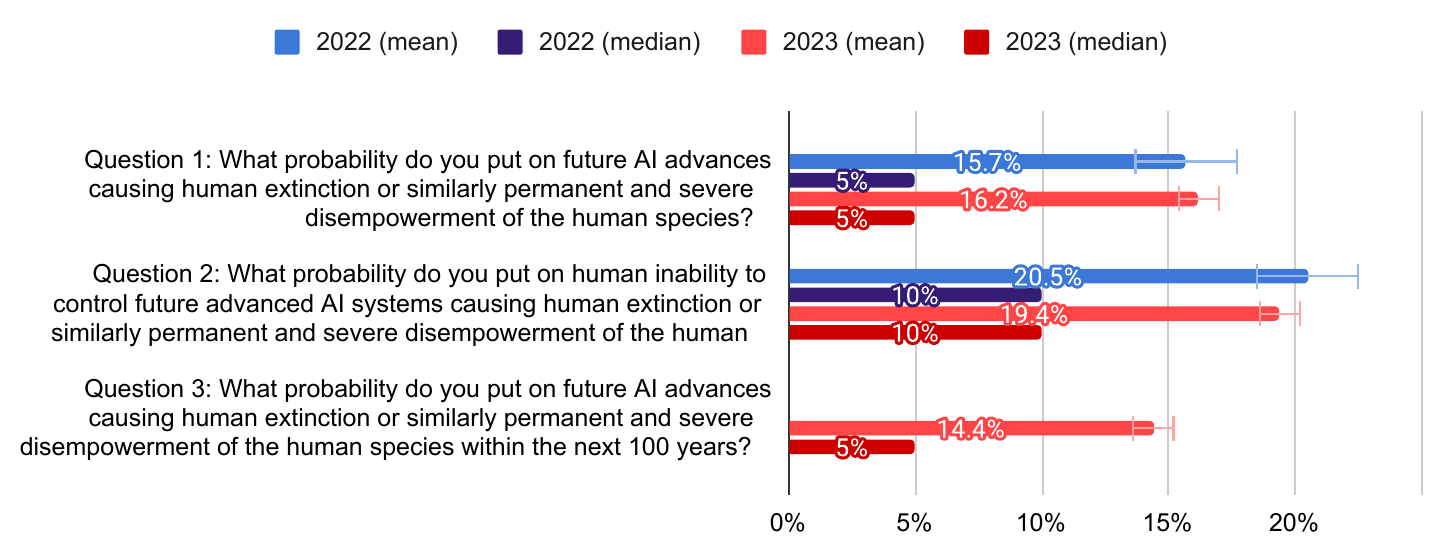}
\caption{\textbf{Mean and median predictions to three questions on human extinction. Error bars indicate the standard error.} (Question 1 n=149 in 2022 and 1321 in 2023. Question 2 n=162 in 2022 and 661 in 2023. Question 3 was asked only in 2023, n=655).}
\label{fig:extinction_predictions}
\end{figure}

\begin{table}
    \centering
    \begin{tabular}{p{2.5in}|p{0.8in}p{1.2in}p{1.2in}}
         & Statistics & 2022 Result & 2023 Result \\
    \hline
    What probability do you put on \textbf{future AI advances} causing human extinction or similarly permanent and severe disempowerment of the human species? & N; Mean (SD); Median (IQR) & 149; 15.7\% (22.1\%); 5\% (19\%) & 1321; 16.2\% (23\%); 5\% (19\%)\\
    What probability do you put on \textbf{human inability to control future advanced AI systems} causing human extinction or similarly permanent and severe disempowerment of the human species? & N; Mean (SD); Median (IQR) & 162; 20.5\% (26.2\%); 10\% (29\%) & 661; 19.4\% (26\%); 10\% (29\%) \\
    What probability do you put on \textbf{future AI advances} causing human extinction or similarly permanent and severe disempowerment of the human species \textbf{within the next 100 years}?\tablefootnote{Bold font was not present in the original questions; it has been added here to emphasize the differences between wordings. Any individual respondent did not see more than one of these questions.} & N; Mean (SD); Median (IQR) & Not asked & 655; 14.4\% (22.2\%); 5\% (19.9\%)\\
    \hline
    \end{tabular}
    \caption{\textbf{Respondents' estimates in 2022 and 2023 for the probability that AI causes human extinction. For each of the two questions that were asked in both years, the results are very similar.}}
    \label{tab:extinction_estimates}
\end{table}

Answers to these questions were mostly consistent, with medians of 5\% or 10\%. These are also close to answers to the question on general value of long-run impact,\footnote{``Assume for the purpose of this question that HLMI will at some point exist. How positive or negative do you expect the overall impact of this to be on humanity, in the long run? Please answer by saying how probable you find the following kinds of impact, with probabilities adding to 100\%''}, which might suggest the bulk of the ``extremely bad (e.g. human extinction)'' answers to that question is from human extinction or similarly permanent and severe disempowerment of the human species, as opposed to other outcomes that respondents to that question may have had in mind but that would have been less severe.

\begin{figure}
\centering
\includegraphics[width=6in]{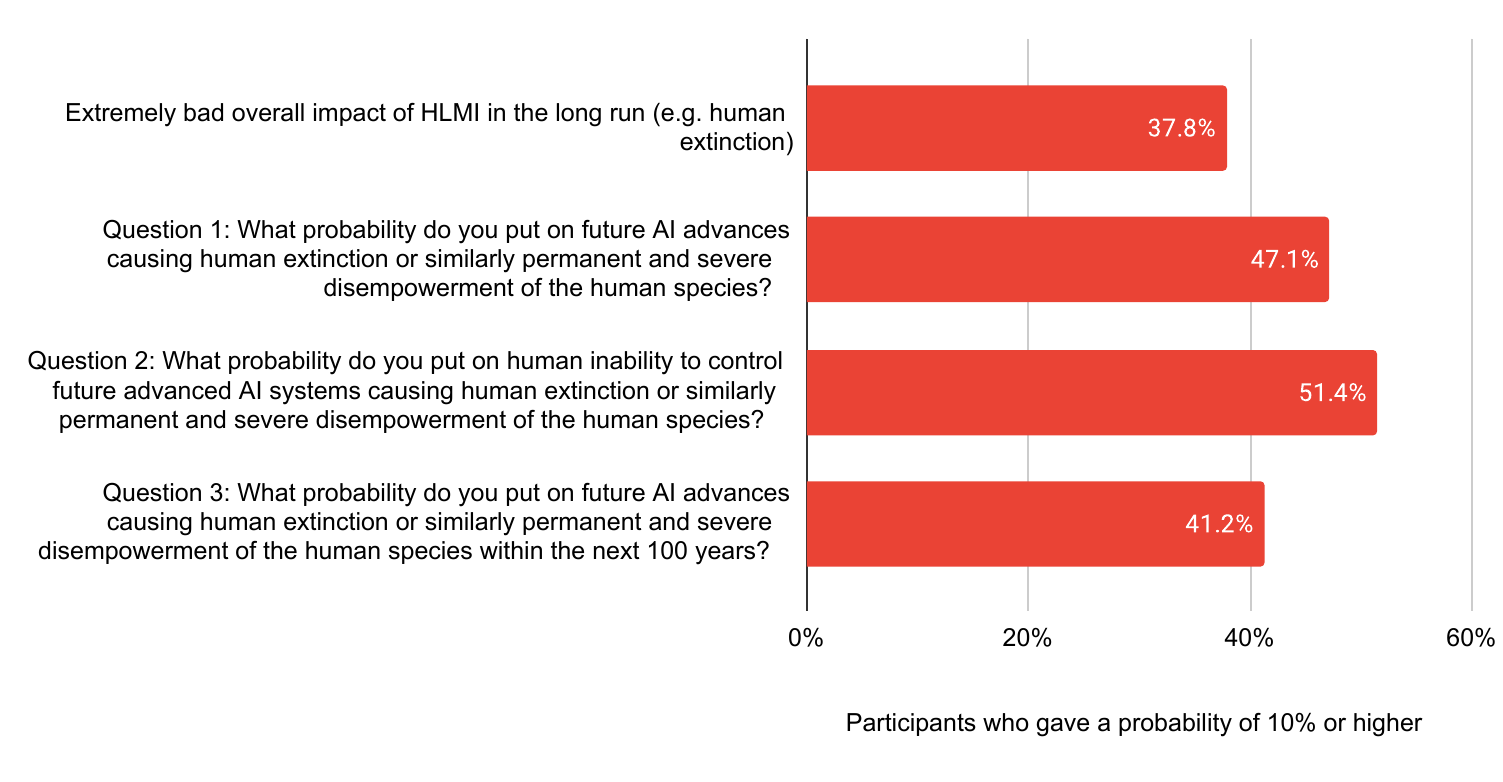}
\caption{\textbf{Percent of participants who gave a probability of 10\% or higher to an extremely bad outcome of HLMI (e.g. human extinction) (see Section~\ref{sec:goodness_of_hlmi}) or to the three questions specifically about human extinction or disempowerment.}}
\label{fig:pdoom10}
\end{figure}

Depending on how we asked, between 41.2\% and 51.4\% of respondents estimated a greater than 10\% chance of human extinction or severe disempowerment (see Figure~\ref{fig:pdoom10}). This is comparable to, but somewhat higher than, the proportion of respondents---38\%---who assigned at least 10\% to ``extremely bad'' outcomes ``(e.g. human extinction)'' in the question asking ``How good or bad for humans will High-Level Machine Intelligence be?'' (See Section~\ref{sec:goodness_of_hlmi})

Appendix~\ref{app:demographics} contains comparisons between results from different demographics on this question.

\subsection{Views on others' concerns about AI}
Respondents were asked a set of ``meta'' questions about their views on others' views (n=671). One question was, ``To what extent do you think people's concerns about future risks from AI are due to misunderstandings of AI research?'' 10.7\% said "Almost entirely.'' 44\% said ``To a large extent.'' 29.1\% said ``Somewhat.'' 14.6\% said ``Not much.'' 1.6\% said ``Hardly at all.'' This may reflect a view amongst AI researchers that the general public misunderstands AI.

\subsection{What rate of AI progress would produce the most optimism?}
We asked participants ``What rate of global AI progress over the next five years would make you feel most optimistic for humanity's future? Assume any change in speed affects all projects equally.'' There was disagreement on whether faster or slower progress would be preferable, though large divergence from the current speed was less popular (Table~\ref{tab:progress_preferences}).

\begin{table}
    \centering
    \begin{tabular}{r|l}
        Answer & Portion of respondents\\
        \hline
        Much slower & 4.8\% \\
        Somewhat slower & 29.9\%  \\
        Current speed & 26.9\% \\
        Somewhat faster & 22.8\% \\
        Much faster & 15.6\% \\
    \end{tabular}
    \caption{\textbf{There was disagreement about whether faster or slower global AI progress over the next five years would be best for humanity's future.}}
    \label{tab:progress_preferences}
\end{table}

\subsection{How much should AI safety research be prioritized?}
We asked some respondents one version of a question about AI research prioritization that matched previous surveys, and we asked other respondents a near-identical question that had been slightly updated. In the old version of the question, we defined AI safety research as follows:

\begin{quote}
Let `AI safety research' include any AI-related research that, rather than being primarily aimed at improving the capabilities of AI systems, is instead primarily aimed at minimizing potential risks of AI systems (beyond what is already accomplished for those goals by increasing AI system capabilities).

Examples of AI safety research might include:
\begin{itemize}
\item Improving the human-interpretability of machine learning algorithms for the purpose of improving the safety and robustness of AI systems, not focused on improving AI capabilities
\item Research on long-term existential risks from AI systems
\item AI-specific formal verification research
\item Policy research about how to maximize the public benefits of AI
\end{itemize}
\end{quote}

The updated question is identical except for the inclusion of this example:
\begin{quote}
\begin{itemize}
\item Developing methodologies to identify, measure, and mitigate biases in AI models to ensure fair and ethical decision-making
\end{itemize}
\end{quote}

\begin{figure}
\centering
\includegraphics[width=6in]{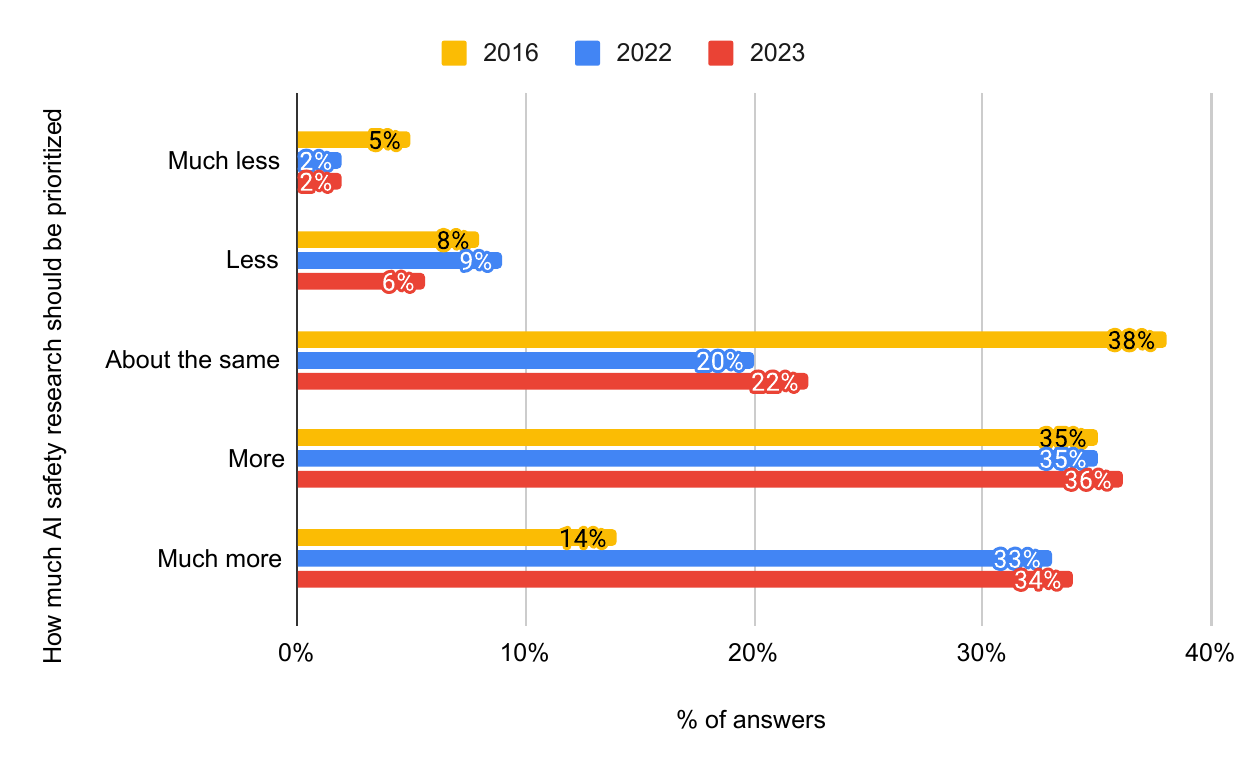}
\caption{\textbf{70\% of respondents thought AI safety research should be prioritized more than it currently is}. Developments since the 2022 survey have not substantially changed the proportion of participants who think AI safety should be prioritized ``more'' ore ``much more''.}
\label{fig:moresafety}
\end{figure}

In both variations, we asked, ``How much should society prioritize AI safety research, relative to how much it is currently prioritized?'' A Welch t-test found that the difference between the two framings was not significant ($t(1327) = -0.58, p=0.564, d=-0.03$), so the results were combined (n=1329). A large majority of respondents thought that AI safety research should be prioritized more than it currently is. The percentage of researchers who thought so increased compared to earlier surveys, but only slightly since 2022. (Figure~\ref{fig:moresafety})

\subsection{How worthy and hard is the alignment problem?}
A second set of AI safety questions was based on Stuart Russell's formulation of the alignment problem \citep{russell2014}. This set of questions began with a summary of Russell's argument---which claims that with advanced AI, ``you get exactly what you ask for, not what you want''---then asked:

\begin{enumerate}
\item Do you think this argument points at an important problem?
\item How valuable is it to work on this problem today, compared to other problems in AI?
\item How hard do you think this problem is, compared to other problems in AI?
\end{enumerate}

\begin{figure}
\centering
\includegraphics[height=3in]{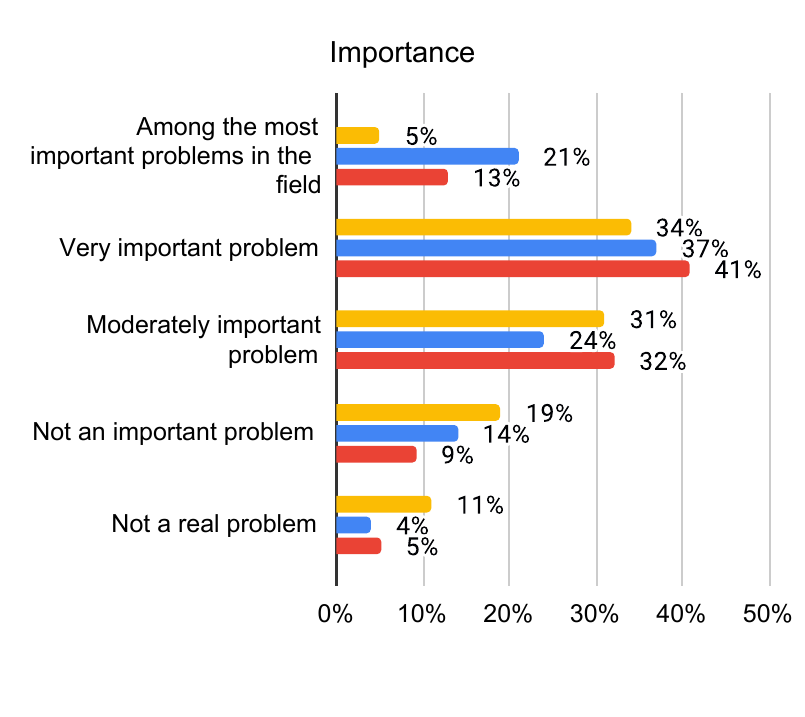}
\includegraphics[height=3in]{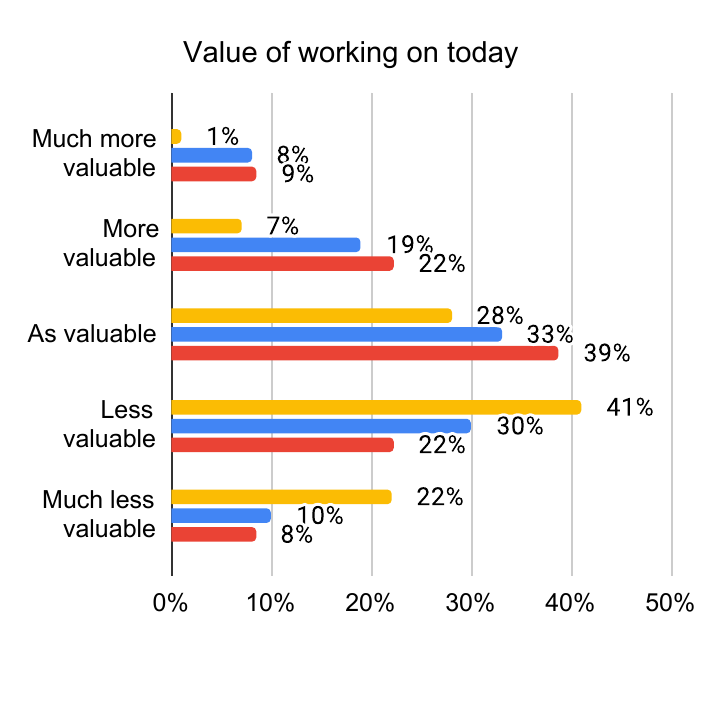}
\includegraphics[height=3in]{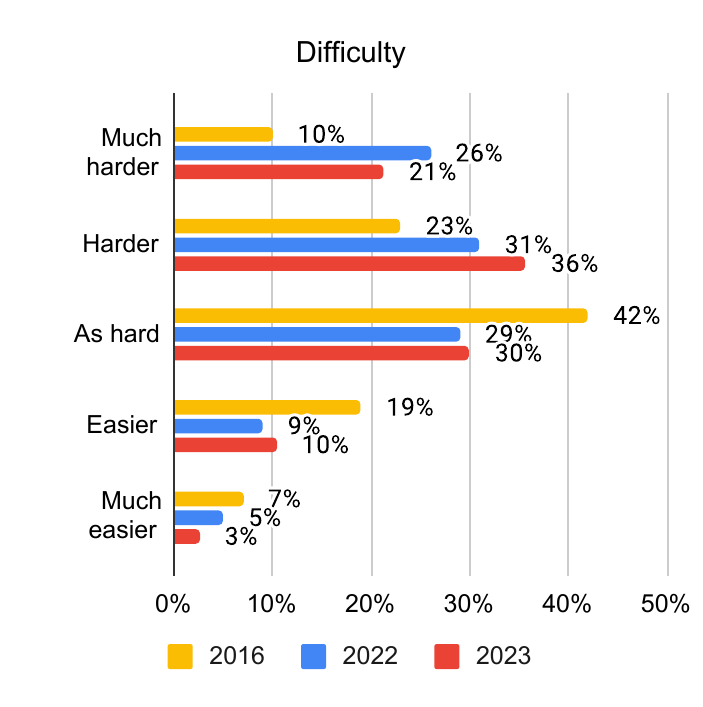}
\caption{\textbf{Attitudes towards Stuart Russell's formulation of the alignment problem.} Participants viewed the alignment problem as important and difficult, but not more valuable to work on than other problems.}
\label{fig:russellalignment}
\end{figure}

The majority of respondents said that the alignment problem is either a ``very important problem'' (41\%) or ``among the most important problems in the field'' (13\%), and the majority said the it is ``harder'' (36\%) or ``much harder'' (21\%) than other problems in AI. However, respondents did not generally think that it is more valuable to work on the alignment problem today than other problems. (Figure~\ref{fig:russellalignment})
\FloatBarrier

\section{Discussion} \label{sec:discussion}
\subsection{Summary of results}
Participants expressed a wide range of views on almost every question: some of the biggest areas of consensus are on how wide-open possibilities for the future appear to be. This uncertainty is striking, but several patterns of opinion are particularly informative.

While the range of views on how long it will take for milestones to be feasible can be broad, this year's survey saw a general shift towards earlier expectations. Over the fourteen months since the last survey \citep{grace2022}, a similar participant pool expected human-level performance 13 to 48 years sooner on average (depending on how the question was phrased), and 21 out of 32 shorter term milestones are now expected earlier.

Another striking pattern is widespread assignment of credence to extremely bad outcomes from AI. As in 2022, a majority of participants considered AI to pose at least a 5\% chance of causing human extinction or similarly permanent and severe disempowerment of the human species, and this result was consistent across four different questions, two assigned to each participant. Across these same questions, between 38\% and 51\% placed at least 10\% chance on advanced AI bringing these extinction-level outcomes (see Figure \ref{fig:pdoom10}). 

In general, there were a wide range of views about expected social consequences of advanced AI, and most people put some weight on both extremely good outcomes and extremely bad outcomes. While the optimistic scenarios reflect AI's potential to revolutionize various aspects of work and life, the pessimistic predictions---particularly those involving extinction-level risks---serve as a stark reminder of the high stakes involved in AI development and deployment.

Concerns were expressed over many topics beyond human extinction: over half of eleven potentially concerning AI scenarios were deemed either ``substantially'' or ``extremely'' concerning by over half of respondents.

\subsection{Caveats and limitations}
\subsubsection{Forecasting is hard, even for experts}
Forecasting is difficult in general, and subject-matter experts have been observed to perform poorly \citep{tetlock2005expert, savage2021strategy}. Our participants' expertise is in AI, and they do not, to our knowledge, have any unusual skill at forecasting in general. 

There are signs in this research and past surveys that these experts are not accurate forecasters across the range of questions we ask. For one thing, on many questions different respondents give very different answers, which limits the number of them who can be close to the truth. Nonetheless, in other contexts, averages from a large set of noisy predictions can still be relatively accurate \citep{surowiecki2004wisdom}, so a question remains as to how informative these aggregate forecasts are.

Another piece of evidence against the accuracy of forecasts is the observation of substantial framing effects (See Sections \ref{sec:hlmi} and \ref{sec:framings}). If seemingly unimportant changes in question framing lead to large changes in responses, this suggests that even aggregate answers to any particular question are not an accurate guide to the answer. In an extreme example in a closely related study, \cite{karger2023forecasting} found college graduates gave answers nearly six orders of magnitude apart when asked in different ways to estimate the size of existential risks from AI: When given example odds of low-probability events, estimates were much lower. A similar effect might apply at some scale to our participants, though their expertise and quantitative training might mitigate it. Participants who had thought more in the past about AI risks seem to give higher numbers, suggesting they are unlikely to give radically lower numbers with further examples of risks (see Table \ref{tab:hlmi_goodness_time_thinking}).

Despite these limitations, AI researchers are well-positioned to contribute to the accuracy of our collective guesses about the future. While unreliable, educated guesses are what we must all rely on, and theirs are informed by expertise in the relevant field. These forecasts should be part of a broader set of evidence from sources such as trends in computer hardware, advancements in AI capabilities, economic analyses, and insights from forecasting experts. However, AI researchers' familiarity with the relevant technology, and experience with the dynamics of its progress, make them among the best-positioned to make informative educated guesses.

\subsubsection{Participation}

The survey was taken by 15\% of those we contacted. This appears to be within the typical range for a large expert survey. Based on an analysis by \cite{hamilton2003online}, the median response rate across 199 surveys was 26\%, and larger invitation lists tend to yield lower response rates: surveys sent to over \numprint{20000} people, like ours, are expected to have a response rate in the range of 10\%. In specialized samples, such as scientists, lower response rates are common due to sampling challenges, as exemplified by \cite{american2009survey}, which achieved a 15.7\% response rate in a survey of life scientists.

As with any survey, our results could be skewed by participation bias, if participants had systematically different opinions than those who chose not to participate. We sought to minimize this effect by aiming to maximize response rate, and by limiting cues about the survey content available before opting to take the survey. We looked for evidence of response bias at the survey level and question level for some questions, and did not find any that would affect the results to a large extent (see Appendix \ref{app:response_bias} for more detail).

\subsubsection{Change in sample from 2022} \label{sec:samplechange}

The two past editions of ESPAI have only surveyed researchers at NeurIPS and ICML, whereas this year we also contacted researchers who published in ICLR, AAAI, IJCAI, and JMLR. This could make comparison between survey years less meaningful, if the populations have different opinions. However where we checked, the subset of 2023 participants who published in NeurIPS or ICML specifically appeared to have very similar opinions to the full sample.

\section{Methods} \label{sec:methods}
In October 2023, we distributed a survey of perspectives about the future of AI to people who had recently published at one of six top-tier AI venues. The questions focused on topics such as the timing of AI progress, the future impact of AI, and AI safety. The survey and its implementation were approved by the Ethics Committee of the University of Bonn (248/23-EP). The survey and its analysis were preregistered \citep{osf_prereg2023}.

\subsection{Survey questions}

Most questions were identical to those asked previously in surveys conducted in 2016 \citep{grace2018} and 2022 \citep{grace2022}. For each of two questions, we added a new version with modified phrasing and randomly assigned participants either the version of the question asked in \cite{grace2022} or the new version. We also added several new questions which were not based on questions from \cite{grace2018} or \cite{grace2022}. While designing new questions, we tested them in a series of interviews with AI researchers and students. The full survey is available on the AI Impacts Wiki (Link in Appendix~\ref{app:data_links}).

As in \cite{grace2018} and \cite{grace2022}, most questions were randomly assigned to only a subset of participants, in order to keep the number of questions for each participant low. The survey was hosted on the Qualtrics survey platform. 

\begin{figure}
\centering
\includegraphics[width=5.5in]{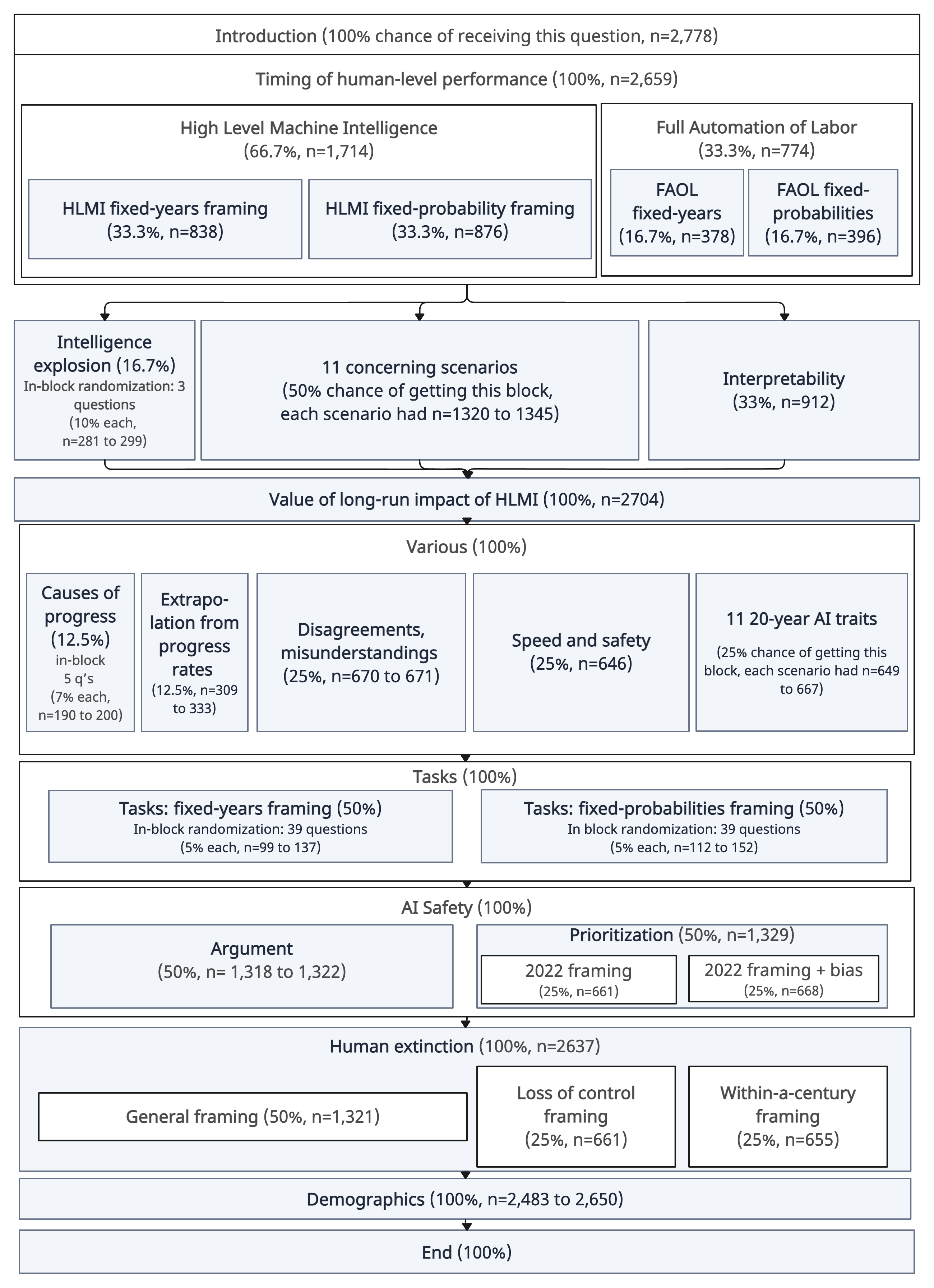}
\caption{\textbf{Individual components of the survey and their randomization.} Horizontal divisions represent participants being randomly split between questions (parenthetical percentages also give the fraction of participants receiving each question block). There was further randomization (not shown), within some question blocks.}
\label{fig:flowdiagram}
\end{figure}

For a full description of the survey’s flow, see the diagram in Figure~\ref{fig:flowdiagram}.

\subsection{Randomization}

Some questions were available in the two types of framing we call the ``fixed-years framing'' and the ``fixed-probabilities framing.'' In the ``fixed-probabilities framing,'' we asked respondents how many years until they thought each AI task would be feasible with a small chance (10\%), an even chance (50\%), and a high chance (90\%). In the ``fixed-years framing,'' we asked respondents how likely they thought it was that each AI task would be feasible within the next 10 years, 20 years and 50 years (or 40 years\footnote{The HLMI-framing used 40 years instead of 50 years. This was done to keep the survey consistent with the previous surveys, where this discrepancy was introduced by mistake.
}). The questions available in these two framings were

\begin{itemize}
\item those asking when narrow AI tasks would become feasible
\item that asking when human-level machine intelligence (HLMI) would become feasible.
\item those asking when occupations would be automated
\end{itemize}

Respondents were randomly allocated to either the ``fixed-years framing'' or the ``fixed-probabilities framing'' (allocation ratio: 1:1) and then received all the questions above using the same framing.

\subsection{Recruitment}

We recruited participants who published in 2022 at any of six top-tier AI venues: the Conference on Neural Information Processing Systems (NeurIPS), the International Conference on Machine Learning (ICML), the International Conference on Learning Representations (ICLR), The AAAI Conference on Artificial Intelligence (AAAI), The Journal of Machine Learning Research (JMLR), and the International Joint Conference on Artificial Intelligence (IJCAI). Compared to the 2022 survey \citep{grace2022}, which was distributed to a randomly-selected half of the researchers who published in 2021 at NeurIPS or ICML, the 2023 survey was distributed to more than three times as many recipients, and to recipients from a wider range of AI specialties.

We collected approximately \numprint{21800} names from publications, then searched for matching emails in those publications, other AI-related publications, and elsewhere. We found email addresses for \numprint{20066} (92\%) of collected names. Using the Qualtrics survey platform, we sent a pre-survey announcement to all of the collected emails on either October 5\textsuperscript{th} (2006 pilot surveys) or October 10\textsuperscript{th}, 2023.

\subsection{Fielding}

Data was collected using an online survey, which was conducted through the Qualtrics survey platform, and delivered via private link in an email.

To investigate the effect of incentives on the response rate, 5\% of the collected emails were assigned to a pilot group to be offered a reward of \$50 reward or an equivalent charitable donation,\footnote{Ultimately participants were also generally offered a further choice of equivalent gift-cards. Some participants were not permitted by the payment service to receive rewards, and we noted in the invitation that in some countries only charitable donations would be available.} and another 5\% to a pilot group to not be offered a reward. On October 11 2023, invitations to complete the survey were sent to the pilot groups. Because the paid pilot survey group had substantially higher response rates, we decided to offer the reward to all participants (including those who had been in the unpaid group). We used the third-party payment service BHN for sending rewards to the bulk of participants.

We sent invitations to the remainder of the collected emails on October 15, 2023 and sent reminders to pilot recipients on the 13\textsuperscript{th}, the 18\textsuperscript{th}, the 20\textsuperscript{th}, the 22\textsuperscript{nd} and the 23\textsuperscript{rd} and to all other participants on the 17\textsuperscript{th}, the 20\textsuperscript{th}, the 22\textsuperscript{nd}, and the 23\textsuperscript{rd}.

When referring to the survey in the emails, we described the survey vaguely to avoid participation bias. Recipients were informed that the results of the survey would be anonymized. For an example of a typical invitation letter, see Appendix~\ref{app:data_links}.

The survey remained open until October 24, 2023. Out of the \numprint{20066} emails we contacted, \numprint{1607} (8\%) bounced or failed, leaving \numprint{18459} functioning email addresses. We received \numprint{2778} responses, for a response rate of 15\%. 95\% of these responses were deemed `finished' by Qualtrics. Because participants received randomized subsets of the questions, the number of responses is far less than \numprint{2778} for most individual questions.

\subsection{Data preparation}

Edits were made to the raw data before analysis in the hope of preserving its intended meaning. For example, we observed cases where participants gave non-numerical answers to numerical questions or reported probabilities for an event having happened that decreased in time. These were deemed to be errors. If a participants’ answers to a fixed-years or fixed-probability question involved probabilities that decreased in time, these answers were removed. Additionally, for participants who answered with decreasing probabilities for at least one question, we removed answers to fixed time questions for which the probabilities summed to 100\%. In total, 196 participants gave answers with decreasing probabilities, and 7 of these users gave answers summing to 100\% for fixed time questions.

Some participants gave answers to fixed-probabilities questions which identified a particular year, rather than a number of years in the future. For example, 2033 instead of 10. To address this, we subtracted 2023 from answers to fixed-years questions greater than 2000 and less than 3000.

Non-numerical answers to numerical questions were removed, except when an unambiguous interpretation was possible. For example, ``10\%'' was changed to 10 and ``<20'' was changed to 20. Some participants entered non-numerical answers to express the view that a probability would never reach a particular value or would reach it at a point infinitely far in the future (for example, ``infinity'' or ``never''). In these cases, the number of years was set to \numprint{100000000}. Our analysis was insensitive to the value of this upper bound.

Data cleaning, analysis, and figure-creation was performed using R statistical software, version 4.3.1, SPSS (Version 29), Google Sheets, and Creately.

\section{Acknowledgments} \label{sec:acknowledgments}
Many thanks for help with this research to Rebecca Ward-Diorio, Jeffrey Heninger, John Salvatier, Nate Silver, Jimmy Rintjema, Joseph Carlsmith, Justis Mills, Will MacAskill, Zach Stein-Perlman, Shakeel Hashim, Mike Levine, Lucius Caviola, Eli Rose, Nathan Young, Michelle Hutchinson, Arden Koehler, Isabel Juniewicz, Ajeya Cotra, Josh Kalla, Niki Howe, Seb Farquar, Nuño Sempere, Naomi Saphra, Sören Mindermann, David Gros, Frederic Arnold, Max Tegmark, Jaan Tallinn, Shahar Avin, Alexander Berger, Cate Hall, Howie Lempel, Jaime Sevilla, Daniel Kokotajlo, Jacob Hilton, James Aung, Ryan Greenblatt, Dan Hendrycks, Alex Tamkin, Vael Gates, Yonadav Shavit, and others. 

We would also like to thank all participants in the survey. The following participants (listed in a random order) agreed to be recognized by name.

\scriptsize
Kun Wu, Dazhong Rong, Achille Thin, Peng Liu, Theo Matricon, Sriram Ganapathi Subramanian, Babak Rahmani, Gustavo De Rosa, Changwoon Choi, Emanuele Sansone, Di Chen, Sheng Guo, Mansooreh Karami, Hansin Ahuja, Leo Tappe, Saed Rezayi, Stephan Waldchen, Gleb Polevoy, Shubhanshu Shekhar, Yingheng Wang, Wenjie Qu, Murat Kocaoglu, Shiyan Jiang, Gregory Rogez, Zhenmei Shi, Qinglin Zhang, Neeraj Wagh, Florian Golemo, Youngeun Nam, Washim Uddin Mondal, Linlin Yang, Eric Zhao, Shuming Kong, Woojun Kim, Franz Pernkopf, Zan Wang, Charles Sutton, Shiri Alouf-Heffetz, Jan Corazza, Iason Gabriel, Shentao Yang, Takuya Ito, Guangming Yao, Aniket Bhatkhande, Jian Liu, Bhavya Kailkhura, Tim G J Rudner, Dachun Sun, Kamil Tagowski, Francesco Ricca, Kaizhao Liang, Gokul Swamy, Sandra Kiefer, Till Richter, Yulong Liu, Selina Meyer, Junyoung Park, Taiki Todo, Yash Chandak, Mark Hamilton, Bo Liu, Zhoujian Sun, Amitoz Azad, Jill-Jenn Vie, Akanksha Saran, Jiafu Wei, Jason Fries, Cheng Wen, Yonggan Fu, Hui Shi, Diego Calvanese, Ramazan Gokberk Cinbis, Shenyang Huang, Yifan Peng, Dahyun Kim, Haochuan Cui, James Requeima, Silpa Vadakkeeveetil Sreelatha, Haibo Chen, Mark Muller, Philipp Holl, Xiong-Hui Chen, Sean B Holden, Rylan Schaeffer, Da-Shan Shiu, Yangze Zhou, Juhan Bae, Robert Martin, Byeonghu Na, Edward Choi, Nathan Wycoff, Joseph Kwon, Michelle Sweering, Aaditya Prakash, Cuiying Gao, Tanuj Agarwal, David Evans, Maximilian Schmidhuber, Kai Yi, Chia-Yi Hsu, Zafeirios Fountas, Rei Kawakami, Brian Ichter, Shikun Li, Jishnu Ray Chowdhury, Ou Wu, Zhaomin Wu, Hung-Nghiep Tran, Benedikt Holtgen, Takuya Hiraoka, Tong Liu, Mikhail Yurochkin, Yanzhi Chen, Julia Niebling, Junting Pan, Yunseok Lee, Pedro J Goncalves, Michael Ryoo, Xinning Chen, Xiongye Xiao, Dongzhan Zhou, Jialun Peng, Yuhui Zhang, Fan Wu, Rachid Riad, Sebastian Schellhammer, Shikun Liu, David Kanter, Bruce XB Yu, Moritz K Lehmann, Hao-Shu Fang, Mohak Goyal, Kwanho Park, Sanghamitra Dutta, Yafei Yang, Hadi Hosseini, Celia Chen, Minseon Kim, Max Simchowitz, Ben Tu, Tim Pearce, Salman Asif, Gianluigi Silvestri, P R Kumar, Daniel Iascone, Wenyue Hua, Chris Cameron, Nicolas Zucchet, Lys Sanz Moreta, Pablo Antonio Moreno Casares, Aadirupa Saha, Maurizio Gabbrielli, Liping Liu, Michal Zajac, Kevin Liu, Xiaowei Jia, Daniel Schneider, Frank Hutter, Xingyu Peng, Antonio Montanaro, I Zeki Yalniz, Jiayang Li, Shujian Yu, Jeff Clune, Changjian Shui, Luca Carminati, Luc Rey-Bellet, Xiang Ruan, Paul Gavrikov, Joshua Cooper, Aymeric Dieuleveut, Leon Derczynski, Paolo Fraccaro, Peter Mernyei, Giung Nam, Zijian Zhang, Peter Bevan, Nisarg Shah, Miao Liu, Yupei Liu, Michael Hutchinson, Xiatian Zhu, Rahmatollah Beheshti, Kyle Hsu, Juraj Podrouzek, Jussi Rintanen, Scott Mueller, Harrison Lee, Nicola Gatti, Zijing Ou, Hiromichi Kamata, Lalit Ghule, Weirui Kuang, Qiang Zhang, Lee Spector, Jiayue Wang, Adams Wei Yu, Arber Zela, Jierui Lin, Justin Sybrandt, Toryn Klassen, Xin Liu, Benjamin Doerr, S Sumitra, Xinbo Zhang, Ryan Burnell, Ashvin Nair, Martha White, Florence Regol, Tobias Golling, Tianlong Chen, Maurizio Ferrari Dacrema, Ahmed M Ahmed, Cem Tekin, Kaizhi Zheng, Ming-Hsuan Yang, Rui Ding, Ziyu Shao, Mariia Seleznova, Byoung Chul Ko, Yogesh Kumar, Aditya Kusupati, Emanuele Aiello, Eric Greenwald, Scott Yih, Jyotirmoy Deshmukh, Dongjin Lee, Matteo Tiezzi, Shenao Zhang, Daiheng Gao, Wentao Zhang, Zhen Fang, Wanqian Zhang, Li Ding, Elizabeth Akinyi Ondula, Sanae Lotfi, Hiroyuki Toda, Shang Gao, Onno Kampman, Xianglin Yang, Tomasz Korbak, Xiaowei Wu, Linhao Luo, Xueyan Niu, Xuan Son Nguyen, Jiawei Du, Dorde Zikelic, Xuan Li, Zongyan Han, Anind Dey, Vineet Gundecha, Yunfan Li, Tongxin Li, Dylan R Ashley, Andre Wibisono, Arnold Smeulders, Euijoon Ahn, Jaakko Lehtinen, Haotao Wang, Hidetaka Kamigaito, Daniel Freeman, Fuxiang Zhang, Kanchana Ranasinghe, Shiqing Wu, Pushi Zhang, Samy-Safwan Ayed, Bingyin Zhao, Fredrik Lofman, Hangjie Yuan, Shu Ding, Pabitra Mitra, Jan Peters, Amartya Mitra, Zi Wang, Feng Liang, Hengguan Huang, Dan Hendrycks, Donghwan Lee, Christoffer Riis, Muhammad Firmansyah Kasim, Botao Yu, Maarten Buyl, Daniel Brown, Joseph Pemberton, Ziwei Xu, Haeyong Kang, Ranjie Duan, Abhishek Roy, Wenpeng Zhang, Zhao Song, Guanghui Yu, Giorgia Dellaferrera, Zicheng Liu, Chuanxia Zheng, Saurabh Garg, Yuchang Sun, Chengan He, Thore Graepel, Andrew Kaploun, Fanchen Bu, Yushun Zhang, Zhuo Lu, Tobias Huber, Huayu Chen, Marko Zeman, Lu Qi, Shin'Ya Yamaguchi, Linxuan Song, Dongsheng Chen, Ilyes Batatia, Yuheng Zhang, Zi-Yi Dou, Michael Kampffmeyer, Shaocong Ma, Benjamin Lebrun, Davin Choo, Christoph Bergmeir, Pavel Tokmakov, Zhifeng Zhang, Samar Khanna, Benjamin Rosman, Shubhankar Mohapatra, Bart Selman, Zongqian Wu, Lei Ding, Anastasiia Varava, Lai Tian, Tiancheng Lin, Andrea Loreggia, Eloy Pinol, Philipp Oberdiek, Zhiyu Chen, Yim Register, Guohao Shen, Maggie Makar, Vit Musil, Greg Durrett, Oumar Kaba, Yongxin Chen, Sepehr Sameni, Yuhang Li, Yiheng Tu, Chen Dan, James U Allingham, Yibo Yang, Sheo Yon Jhin, Baris Coskunuzer, Thomas Unterthiner, Enric Boix-Adsera, Nan Wu, James Ostrowski, James Simon, Asadur Chowdury, Binghong Chen, Daniel Neider, Quentin Delfosse, Zeren Shui, Krishna Pillutla, Shangmin Guo, Naiyan Wang, David Johnson, Tirtharaj Dash, Stephan Clemencon, Zhan Ling, Yaman Kumar Singla, Hassam Sheikh, Samuel Showalter, Luckeciano C Melo, Mike Laszkiewicz, Gaurang Sriramanan, Hannes Stark, Luo Mai, Junkang Li, Lukasz Janeczko, Daniel P M De Mello, Derek Lim, Marcel Worring, Julian Bitterwolf, Cheng Jin, Ziwei Wu, Wenqi Zhang, Kun Jing, Ronilo Ragodos, Bart Van Merrienboer, Ruijie Jiang, Minyi Zhao, Thomas Moreau, Abbavaram Gowtham Reddy, Xianyuan Zhan, Daniele Zambon, Kyu-Hwan Jung, Yankai Jiang, Zhuqing Liu, Zhenbo Song, Daniel Widdowson, Vandan Gorade, Jiesong Liu, Tingting Liang, Trung X Pham, Tahseen Rabbani, Cheng Soon Ong, Ailin Deng, Michele Flammini, Fei Zhang, Beihao Xia, Lewis Griffin, Eunseon Seong, Zou Xin, Yongjie Yang, Aditya Mahajan, Ting Li, Runsheng Yu, Aditya Gopalan, Kien Do, Andrew Patterson, Rishi Sonthalia, Yuqi Ren, Diana Borsa, Stanislaw Jastrzebski, Tenghui Li, Le Minh Binh, Jiashuo Liu, Nabeel Seedat, Amine Aboussalah, Simon Woodhead, Dmitrii Zhemchuzhnikov, Samiul Alam, Oliviero Nardi, Wei Xing, Zhuofan Ying, Jeffrey Ryan Willette, Alexander Rudnicky, Jeremiah Birrell, Rudrasis Chakraborty, Luigi Acerbi, Teng Xiao, Zhong Ji, Tomer Galanti, Jesse Engel, Jose Henrique De M Goulart, Lijun Zhang, Feng Zhou, Andrew Zhao, Zichen Zhang, Zhenghua Xu, Rongtao Xu, Felix Petersen, Ammarah Farooq, Joey Tianyi Zhou, Alexander Bergman, Ameya Joshi, Shengyi Huang, Yongtao Wang, Sergei Grudinin, Junjie Hu, Adarsh Kappiyath, Abibulla Atawulla, Alex Rogozhnikov, Lorenzo Perini, Yixuan Su, Renxiong Liu, Zhenyu Zhu, Sai Srinivas Kancheti, Mingyu Kim, Wei Wan, Fabian Latorre, Charles Guille-Escuret, Elvis Nava, Dave De Jonge, Nikola Konstantinov, Renan A Rojas-Gomez, Jun Shern Chan, Mengfan Wang, Lirong Wu, Haoyang Li, Michiel Bakker, Youngjo Lee, Shane Rogers, Zhiyuan Wen, Zaiqiao Meng, Yun Zhu, Yurong Chen, Hans Hao-Hsun Hsu, Elena Stefancova, Ningyu Zhang, Ryo Karakida, Taher Jafferjee, Yongming Rao, Yeonsik Jo, Yoshua Bengio, Taesup Kim, Cesar Ferri, Eirik Lund Flogard, Robert Kaufman, Wonkwang Lee, Stanislaw Szufa, Stephen O Mussmann, Ido Greenberg, Jie Qiao, Matthew D Hoffman, Zoran Tiganj, Akhilan Boopathy, Alessandro Farinelli, Weishun Zhong, Siddharth Nayak, Xiaolei Diao, Yuanbiao Gou, Zhenting Wang, Shu Hu, Augustine Mavor-Parker, David Ruhe, Hongyi Guo, Stephen Casper, Yifei Min, Bernardo Subercaseaux, Michelangelo Conserva, Haoran Xu, Dmitriy Skougarevskiy, Hanhan Zhou, Christopher Mutschler, Naeemullah Khan, Guosheng Hu, Yufei Huang, Ilya Tyagin, Edoardo Cetin, Davis Brown, Enyan Dai, Filippo Furfaro, Kion Fallah, Bonifaz Stuhr, Dylan Richard Muir, Szymon Tworkowski, Wooyong Jung, Xuebin Zhao, Hehuan Ma, Xi Susie Rao, Maxime Cordy, Yuan Shen, Chengshi Zheng, Farnam Mansouri, Felix Draxler, Ajay Mandlekar, Kai Han, Jim Laredo, Shiji Zhou, Haripriya Harikumar, Alistair Letcher, Jack Valmadre, Minghao Liu, Dorota Celinska-Kopczynska, Lingjiong Zhu, Ben Evans, Lucas Monteiro Paes, Yogesh Verma, Pin-Yu Chen, Steven Basart, Xinpeng Liu, Sebastian Gruber, Daoyuan Chen, Aurelie Lozano, Cenk Baykal, Milad Jalali Asadabadi, Muhammad Uzair Khattak, James M Murphy, Tianyu Cui, Pavlo Melnyk, Emanuele Natale, Eli Upfal, Minji Kim, Angjoo Kanazawa, Pavel Izmailov, Tao Song, Bang Liu, Munyque Mittelmann, Tim Roith, Mingtian Zhang, Oluwadamilola Fasina, Michael Felsberg, Sicong Huang, Steven Stenberg Hansen, Pan Xu, Biraj Dahal, Nicklas Hansen, Xikun Zhang, Francesco Belardinelli, Rui Chen, Jianguo Cao, Paulo Rauber, Yangjun Ruan, Xianqiang Lyu, Qiang Sheng, Spencer Peters, Alexey Luchinsky, Jiao Sun, Hanqing Zhu, Guoji Fu, Qingqing Zhao, Zhongyu Huang, Sihyun Yu, Suraj Kothawade, Luofeng Liao, Sergey Nikolenko, Ken Kahn, Qian Peng, Balaraman Ravindran, Panagiotis Karras, Ivan Serina, Giada Pistilli, Hyolim Kang, Hanchen Wang, Wenhao Wu, Peilin Zhou, Mehrtash Harandi, Jiaqi Han, Yin Fang, An Xu, Tom Bewley, Jiayi Wang, Franz Scherr, Zhengliang Liu, Yiqun Liu, Feng Liu, Difei Gao, Sixing Yu, Tsvetomila Mihaylova, Katherine M Collins, Siyi Tang, Zhipeng Liang, Samarth Mishra, Tyler H Mccormick, Bohang Zhang, Bastian Rieck, Yuangang Pan, Jack Zhang, Johannes Treutlein, Roman Levin, Xiangtai Li, Yangfeng Ji, Una-May O'Reilly, Lior Rokach, Yunjiang Jiang, Sung Ju Hwang, Eduardo Dadalto, Bo Pan, Guendalina Righetti, Johannes Jakubik, Yanchao Sun, Hugo Berard, Eric Mitchell, Xuefeng Du, Seung-Bin Kim, Martin Burger, Yu Luo, Peter Macgregor, Nikolaus Howe, Erik Jenner, Florian Jug, Runlin Lei, Yan Zhuang, Wei-Cheng Tseng, Craig Sherstan, Luojun Lin, Kai Yue, Anastasios Angelopoulos, Dzung Phan, Yuqi Bu, Junseok Lee, Safa Alver, Mengchen Zhao, Ali Seyfi, Chunyang Fu, Minghuan Liu, Enrico Scala, Neel Nanda, Shijie Liu, Roger B Grosse, Wen Song, Gautam Shroff, Mingming Gong, Yushun Dong, Yixin Zhang, Zijie Zhang, Ziqi Liu, Xingran Chen, Vincent Josua Hellendoorn, Xiaoqing Tan, Kuan-Lin Chen, Dingfeng Shi, Simon Schierreich, Concetto Spampinato, Rob Brekelmans, Daniel D Lundstrom, Vahid Balazadeh Meresht, Viraj Prabhu, Zuxin Liu, Ronald J Brachman, Yutong Chen, Liang Zeng, Georgios Birmpas, Xu Chen, Ayush Jain, Yuanwei Liu, Lawson Wong, Muyang Li, Junfeng Guo, Rayan Mazouz, Changho Shin, Yunjuan Wang, Charlie Dickens, Chongjian Ge, Zhijiang Guo, Oleg Vasilyev, Jean Barbier, Huifeng Yao, Luca Citi, Loay Mualem, Francesco Antici, Haozhi Qi, Flavio Miguel Varejao, Jiaxuan Wang, Mohammad Azam Khan, Jun Bai, Vaden Masrani, Zhiwen Fan, Ian Stavness, Mingjian Wen, Qian Tao, Luyao Tang, Hanshu Yan, Chi Wang, Ulisses Braga-Neto, Hui-Po Wang, Anurag Arnab, Marc Law, Gabriele Corso, Pratyush Kumar, Shiyan Chen, Zizheng Pan, Yujia Qin, Vaisakh Shaj, Jinfa Huang, Volker Gruhn, Marcus A Brubaker, Marco Favorito, Jiayuan Liu, Kartik Gupta, Zishun Yu, Aodong Li, Manolis Pitsikalis, Ivan F Rodriguez Rodriguez, Victoria Sherratt, Jaekyeom Kim, Felipe Meneguzzi, Chen-Lu Ding, Lionel Riou-Durand, Edward J Hu, Mingjie Sun, Jinheon Baek, Michael Witbrock, Mike Holenderski, Viraj Mehta, Freddie Kalaitzis, Vicenc Gomez, Yu Pan, Devansh Arpit, Deeksha Arya, Zhipeng Bao, Prince Osei Aboagye, Tomasz Odrzygozdz, Dimitar Iliev Dimitrov, Yasmeen Alsaedy, Alistair Muldal, Maksym Andriushchenko, Elena Zheleva, Rupesh Kumar Srivastava, Jinyoung Han, David Gros, Shangbin Feng, Hao-Tsung Yang, Nithin Nagaraj, Jung-Hun Kim, Leon Hetzel, Garrett Kenyon, Orr Paradise, Yuxuan Yi, Jason Xiaotian Dou, Andreas Loukas, Hyungjin Chung, Jin-Duk Park, Eric Zelikman, Sebastian Shenghong Tay, Jiajun Tang, Kincaid Macdonald, Wonbin Kweon, Haoyue Dai, Achraf Azize, Calvin Tsay, Agnieszka Lawrynowicz, Qiyu Kang, Bingxin Zhou, Alena Shilova, Yuxiao Huang, Robert Honig, Sean Heigeartaigh, Xovee Xu, Martin Gjoreski, Federico Berto, Shichao Xu, Zhigao Guo, Qian Pan, Wenjie Qiu, Charles X Ling, Krzysztof Maziarz, Diane Oyen, Luigi Gresele, Yizhen Zheng, Changcheng Tang, Parth Patwa, Tomas Peitl, Anuroop Sriram, Dhruv Madeka, Hang Wang, John Aslanides, Hyeong Kyu Choi, Yongjun He, Joel Rixen, Krzysztof Sornat, Zhihan Gao, Renzhe Xu, Sebastian Stein, Lin Gu, Vianney Perchet, Javier De La Rosa, Luca Ragazzi, John M Dolan, Jinyu Cai, Krishnaram Kenthapadi, Alex Morehead, Tim Coleman, Animesh Garg, Jianhao Yan, Mihai Nica, Kaichun Mo, Maosen Li, Haoyu Chen, Dong-Hee Paek, Johannes Oetsch, Yizhou Zhao, Seon-Ho Lee, Prasanna Parthasarathi, Haotian Bai, Shilong Liu, Zhiqi Huang, Jonathan Brophy, Vincent Stimper, Jinxin Liu, Kanji Uchino, William Harvey, Xu Luo, Ulrich Aivodji, Frederic Piedboeuf, Yang Tang, Joonhyun Jeong, Jeff Calder, Tao Wu, Wenjie Li, Stefano Panzeri, Kunal Pratap Singh, Cynthia Matuszek, Ke Sun, Ching-Yun Ko, Micah Goldblum, Jiayu Chen, Geon Lee, Zhiyuan Fang, Chi-Ming Chung, Zixia Jia, Peng Ye, Jedrzej Potoniec, Yash Akhauri, Chung-Yeon Lee, Asif Khan, Thomas Meyer, Pablo Barcelo, Salman Khan, Anna Ivanova, Dzmitry Tsishkou, Gabriel Kreiman, Fang-Lue Zhang, Jingkai Zhou, Minchul Shin, Ross M Clarke, Qizhang Feng, Fuqi Jia, Nicolas Troquard, Desik Rengarajan, Sebastian Dalleiger, Selena Zihan Ling, Phil Pope, Beier Zhu, Murat Kantarcioglu, Dianwen Ng, Chunwei Ma, Yi Wu, Qihang Lin, Shaunak Srivastava, Zhuo Li, Muxuan Liang, Tsuyoshi Ide, Kenny Young, Umang Bhatt, Nicolas Lanzetti, Nayeon Lee, Mengchu Li, Joan Puigcerver, Lingjie Mei, Wooseok Shin, Xiaokang Chen, Junpei Komiyama, Dhruv Batra, Melanie Sclar, Marta Kryven, Yihan Zhang, Matthias Weissenbacher, Diego Granziol, Saurabh Sihag, Batu Ozturkler, Eduard Hovy, Yuning You, Ruizhong Qiu, Jiaru Zhang, Anshuman Chhabra, Didier Chetelat, Joey Bose, Raphael Poulain, Ben Armstrong, Xin Wen, Youngmoon Lee, Daniel Karl I Weidele, Tom Dhaene, Tom Drummond, Sangwoo Mo, Anh Tong, Haoliang Li, Xu Cao, Gergely Flamich, Yonggang Zhang, Yao-Yuan Yang, Hiroki Furuta, Shitao Tang, Huang Bojun, Jianing Zhu, Hongyao Tang, Benjamin Guedj, Cade Gordon, Tao He, Daria Stepanova, Li Chen, Timothy Hospedales, Gaetan Marceau Caron, Mancheng Meng, Fabien Gandon, Son Tran, Paterne Gahungu, Elliot Paquette, Lauro Langosco Di Langosco, Zhuoyue Lyu, Zhixiang Wei, Gaurav Pandey, Ziheng Duan, Elisa Fromont, Stefan Heidekruger, Borja G Leon, Meng Liu, Mira Welner, Gang Yan, Giovanni Chierchia, Rasmus Pagh, Thomas Pierrot, Patryk Wielopolski, Ivan Marisca, Matthias Samwald, Yangfu Zhu, Zhuo Chen, Yining Ma, Joseph Early, Luziwei Leng, Laura Wynter, Louis Bethune, Vincent Mai, Matthew I Swindall, Julia Grabinski, John Burden, Ankit Vani, Guillaume Leclerc, Tianying Ji, Soroush Vosoughi, Fermin Travi, Hai Zhao, Xuan Di, Ruofan Liang, Gregor Bachmann, Pradeep Ravikumar, Taoan Huang, Weina Jin, Jiayuan Mao, Deepak Ramachandran, Xiangzhe Kong, Ziyao Huang, Yuko Ishiwaka, Giancarlo Kerg, Joanna Hong, Ziheng Zeng, Seong Joon Oh, Renrui Zhang, Abhishek Gupta, Cole M Wyeth, Jinguo Zhu, Peter Kairouz, Jian Zhao, Soren Mindermann, Bastien Maubert, Longhui Yu, Sujan Dutta, Balaji Vasan Srinivasan, Michael Littman, Mariusz Oszust, Edouard Yvinec, Nishanth Dikkala, Kate Sanders, Hongzhi Wu, Emir Ceyani, Geraud Nangue Tasse, Georgios Georgakis, Zhenfeng He, Hyunwoo J Kim, Danyal Saeed, Bhumika Mistry, Sirisha Rambhatla, Canzhe Zhao, Zulong Chen, Louis Tiao, Marco Console, Tahar Allouche, Seungyong Moon, Daniel Mackinlay, Jing Liu, Kuan Yew Leong, Qingbiao Li, Anwen Hu, Arman Zharmagambetov, Jiawei Shao, Andrew Li, Luke E Richards, Aurelien Garivier, Danda Pani Paudel, Riccardo Marin, Chih-Kuan Yeh, Chao Qin, Ye Zhu, Yinglin Duan, I-Hsiang Chen, Kartik Chandra, Yi Zeng, John E Laird, Leonardo Amado, Luis Fernando D'Haro, Zeyu Qin, Santu Rana, Jiajun Fan, Joshua Welch, Adam Prugel-Bennett, Tristan Bepler, Tai Sing Lee, Georg Ostrovski, Mohammed Haroon Dupty, Niclas Boehmer, Yinkai Wang, Fabian Paischer, Sihan Liu, Drew Linsley, Wuwei Lin, Ziyi Chen, Carol Long, Seongku Kang, Haoyuan Chen, Danyang Tu, Vaibhav V Unhelkar, Ivan Orsolic, Ivan Donadello, Jacob Euler-Rolle, Mohan Zhang, Matej Racinsky, Jiaqing Liang, Ravinder Bhattoo, Javier Segovia-Aguas, Yifan Yang, Isabelle Augenstein, David Lindner, Keane Lucas, Huan Zhang, Hyung Ju Suh, Nikolaos Tsilivis, Pandeng Li, Evgeny S Saveliev, Wendi Li, Shengchao Liu, Pierre Tassel, Avi Mohan, Jinjin Gu, Lapo Faggi, Zhen Zhu, Vihang Patil, Giuseppe Vietri, Morten Goodwin, Zizheng Yan, Erik Daxberger, Yaqian Zhang, Michael Niemeyer, Mauricio Gruppi, Yipeng Qin, Amir Rahmati, Pietro Mazzaglia, Yiwen Kou, Karan Singhal, Fatih Erdem Kizilkaya, Xiangyi Chen, Haoxuan Qu, Lei Jiao, Emanuele Marconato, Zhen-Duo Chen, Cecilia G Morales, James P Delgrande, Zhengyuan Zhou, Vasu Singla, Qian Chen, Guojun Xiong, Peter Stone, Blake Richards, James M Murray, Yu Huang, Tim Verbelen, Edward De Brouwer, Nur Muhammad Shafiullah, Zhenke Wu, Konstantinos Voudouris, Nuo Xu, Youngkyoung Kim, Teemu Roos, Mohammad Reza Belbasi, Wenda Li, Shengyang Sun, Olivier Graffeuille, Marc Dymetman, Yunzhe Li, Markus Hiller, Zhenhailong Wang, Filip Cano Cordoba, Xingjian Du, Zongzhang Zhang, Sandeep Silwal, Mengping Yang, Fabricio Vasselai, Aaron Defazio, Samuel Yang-Zhao, Roberto Posenato, Alexis Joly, Steven Yin, Pedro Bizarro, Chujie Zheng, Subrata Mitra, Devin Garg, Jun Fang, Emanuele Guidotti, Roy Fox, Weiming Liu, Steve Peterson, Harsh M Patel, Dan Navon, Romain Mueller, Rupert Freeman, Ilya Chugunov, Spyridon Samothrakis, Andrea Cini, Kumar Kshitij Patel, Shadi Albarqouni, Xian Yeow Lee, Yichong Leng, Junkyu Lee, Hezhen Hu, Antoine Bosselut, Udayan Khurana, Wee Sun Lee, Dino Sejdinovic, Jiancheng Yang, Ana Bazzan, Mario A T Figueiredo, Eunggu Yun, Marco Roveri, Frederik Schubert, Liushuai Shi, Misoo Kim, Ziyang Song, Xiaoyang Wu, Fabrizio Silvestri, Alan Yuille, Xizewen Han, Mykel J Kochenderfer, Keshav Bhandari, Fan Lyu, Sehyun Hwang, Sheng Li, Yuzheng Hu, Todd W Neller, Wanhua Li, Philipp Tholke, Jason Hartford, Sunwoo Lee, Weiqiang Zheng, Yi Cheng, Jithin Jagannath, Yangrui Chen, Jie Yang, Anders Karlsson, Hongsheng Hu, Zhenbin Wang, Xiaohao Xu, Carlos Escolano, Khai Nguyen, Zhen Dong, Helge Spieker, Honghao Wei, Mingzhi Dong, Zhiyu Lin, Ilsang Ohn, Jiaxi Ying, Namgyu Ho, Stephan Hoyer, Juan Esteban Kamienkowski, Haoran Liu, Yinan Feng, Lingfeng Sun, Mingyuan Zhang, Songhua Wu, Zhichao Wang, Stefano Leonardi, Jose Hernandez-Orallo, Chenyang Wu, He Sun, Adrian Feiguin, Chaoyue Liu, Xiaohua Wu, Yiren Zhao, Nicholas Roberts, Zhaowei Zhu, Fei Huang, Gemine Vivone, Scott Fujimoto, Kerstin Bach, Samuel I Holt, Ashok Narendranath, Elliot Nelson, Zhenyu Yang, Safa C Medin, Mark Crowley, Liyuan Wang, Maria Hedblom, Alfio Ferrara, Pratik Vaishnavi, Ernesto Evgeniy Sanches Shayda, Mireille El Gheche, Ariel Kwiatkowski, Oliver Kiss, Bingchen Zhao, Eddie Ubri, Mert Yuksekgonul, Minyang Hu, Jessica Shi, Changan Niu, Yoav Wald, Maxwell Xu, Nadjib Lazaar, Liangli Zhen, Yaochen Xie, Erdem Biyik, Tian Zhou, Ye Yuan.

\normalsize

\section{Funding} \label{sec:funding}
This research project was funded by Open Philanthropy, 182 Howard Street \#225. San Francisco, CA 94105, USA.

\section{Additional Information} \label{sec:additional_info}
\textbf{Correspondence and requests for materials} should be addressed to Katja Grace.

\textbf{Reprints and permissions requests} should be directed to Katja Grace.

\nocite{*}
\bibliography{references}

\appendix
\newpage
\section{Results Comparing Different Demographics} \label{app:demographics}
\subsection{How soon will High-level Machine Intelligence be feasible?}

\FloatBarrier
\begin{figure}[ht]
\centering
\includegraphics[width=2.5in]{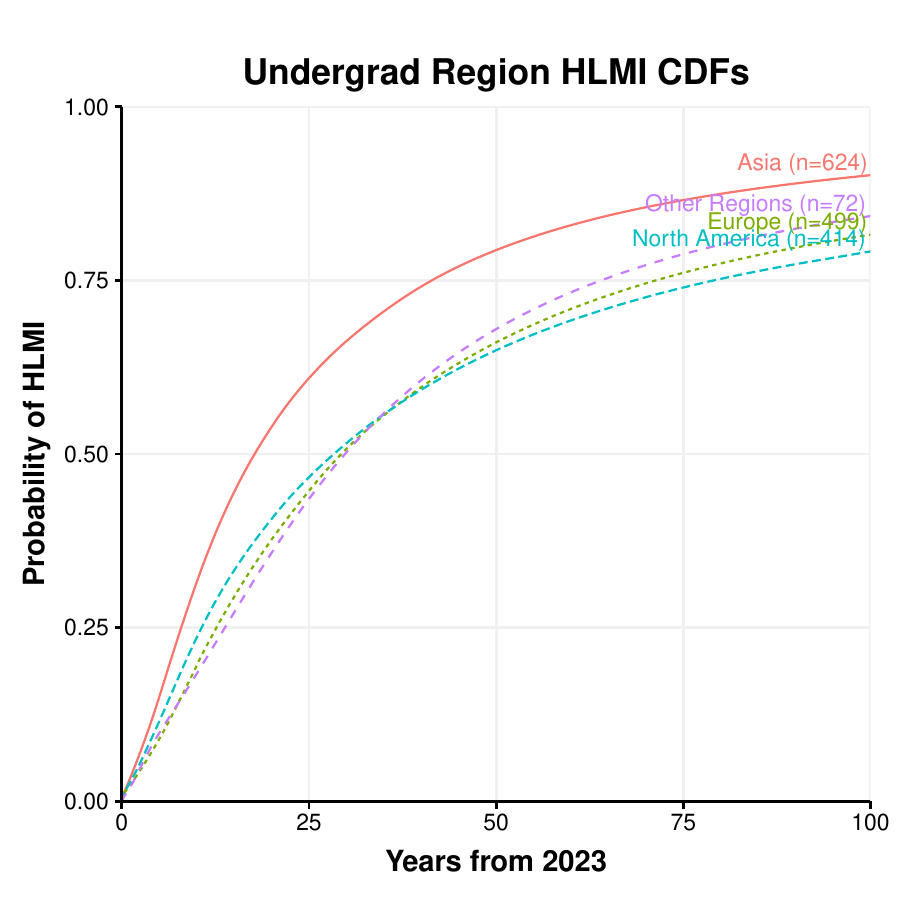}
\includegraphics[width=2.5in]{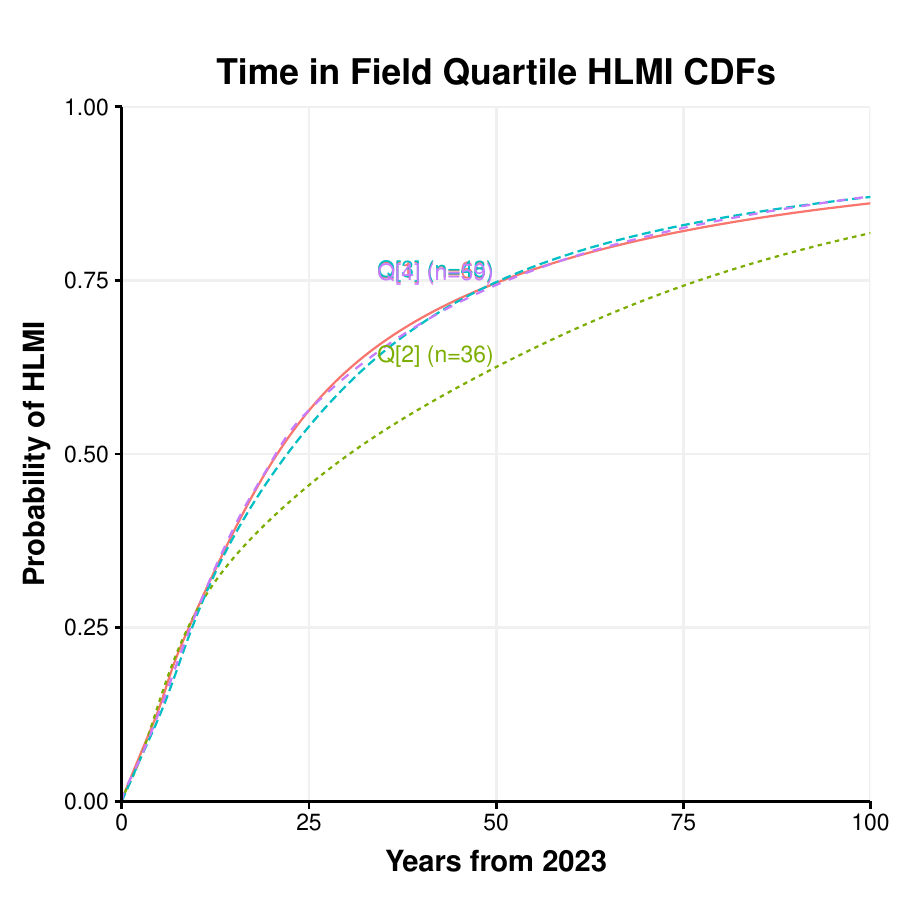}
\includegraphics[width=2.5in]{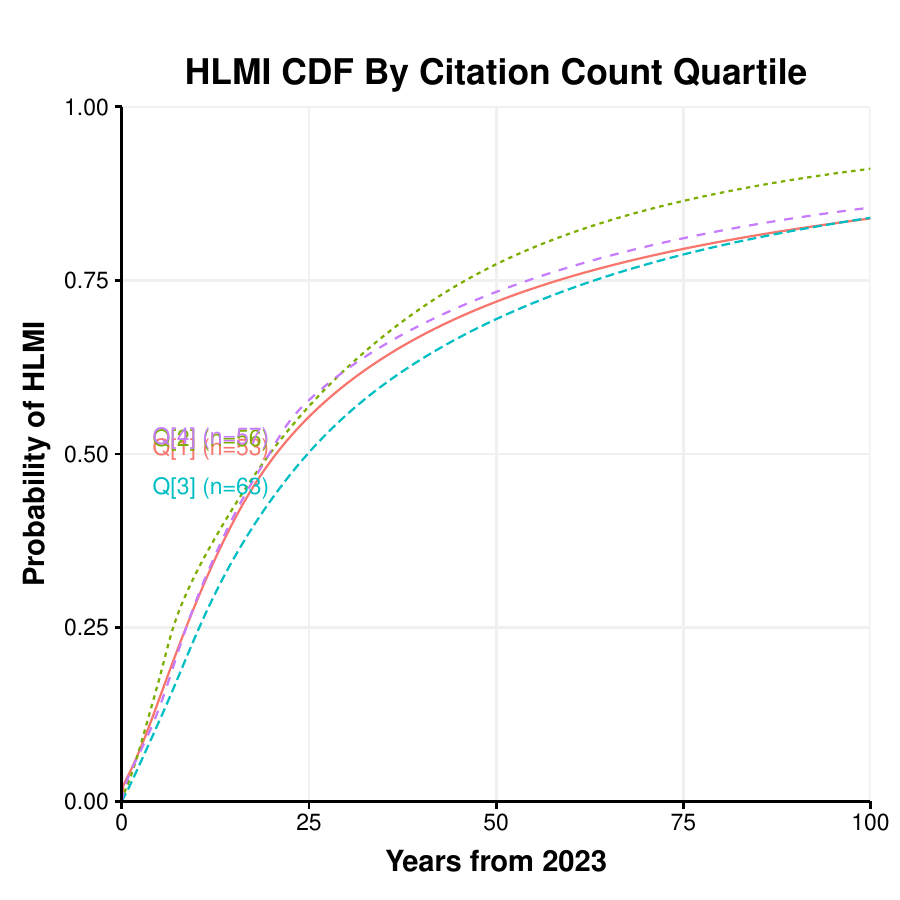}
\label{fig:undergrad_region_hlmi}
\label{fig:time_in_field_hlmi}
\label{fig:citation_count_hlmi}
\caption{\textbf{Aggregate forecasts for time until HLMI were shorter for participants whose region of undergraduate study was Asia. Time in field did not have a significant effect on forecasts for time until HLMI. Citation count did not have a significant effect on forecasts for time until HLMI.}}
\end{figure}

\FloatBarrier
\subsection{How good or bad for humans will HLMI be?}
\subsubsection{Amount of thinking about the issue}

We asked all respondents how much thought they have given to the social impacts of smarter-than-human machines.

\begin{table}[ht]
    \centering
    \begin{tabular}{p{3.5in}|p{1.25in}p{1.25in}}
         & ``Very little'' or ``a little'' thought (n = 621) & ``A lot'' or ``a great deal'' of thought (n = 1113) \\
    \hline
        Percentage of these respondents who thought that the probability of an extremely \textbf{good} outcome was greater than 10\%  & 68\% & 41\%\\
        Percentage of these respondents who thought that the probability of an extremely \textbf{bad} outcome was greater than 10\% & 34\% & 71\% \\
     \hline
    \end{tabular}
    \caption{\textbf{People who had thought more about the social impacts of smarter-than-human machines were substantially less likely to give more than than 10\% credence to extremely optimistic outcomes, and substantially more likely to give extremely negative outcomes more than 10\% credence.}}
    \label{tab:hlmi_goodness_time_thinking}
\end{table}

It is not clear how to interpret the results of this question. Specifically, while thinking more about a topic presumably improves predictions about it, people who think a lot may do so because they are concerned, so the association could also be due to this selection effect.

\FloatBarrier
\subsubsection{Academia vs industry, undergraduate region, and amount of thought}

These are the demographic differences within the answers to the question about how good or bad HLMI will be overall.

\begin{table}[ht]
    \label{tab:demographic_outcome_probs}
    \centering
    \resizebox{6.5in}{!}{
    \begin{tabular}{r|ccccc}
        & Extremely good & On balance good & Neutral & On balance bad & Extremely bad \\
        \hline
        Industry & 12.75 & 30 & 20 & 15 & 5 \\
        Academia & 10 & 25 & 20 & 15 & 5 \\
        Asia & 15 & 30 & 20 & 10 & 5 \\
        North America & 10 & 25 & 20 & 15 & 5 \\
        Other location & 10 & 25 & 20 & 17.5 & 5 \\
        Having thought more & 15 & 25 & 20 & 15 & 5 \\
        Having thought less & 10 & 30 & 20 & 12 & 5 \\
        \hline
    \end{tabular}}
    \caption{\textbf{Median probabilities given to each outcome by each demographic group.}}
\end{table}

\FloatBarrier
\subsection{How likely is AI to cause human extinction?}

Whether participants worked in academia or industry did not affect median responses much—both had a median of 5\%. The region respondents graduated from affected responses somewhat: the Asian median was 10\%, while North American and European medians were 5\%. Having thought more (either ``a lot'' or ``a great deal'') about the question was associated with a median of 9\%, while having thought ``little'' or ``very little'' was associated with a median of 5\%.

The similarity of answers across several slightly different questions, across this and previous surveys, and across participants in academia and industry and different geographic regions appears to be robust evidence that the majority of AI researchers think there is a nontrivial risk of extinction or similar catastrophes due to AI.

\newpage
\section{Supplementary Figures} \label{app:supplementary_figures}
\FloatBarrier
\subsection{How soon will human-level performance on all tasks or occupations be feasible?}

\hspace{0pt}
\vfill
\begin{figure}[h]
\centering
\includegraphics[width=4.2in]{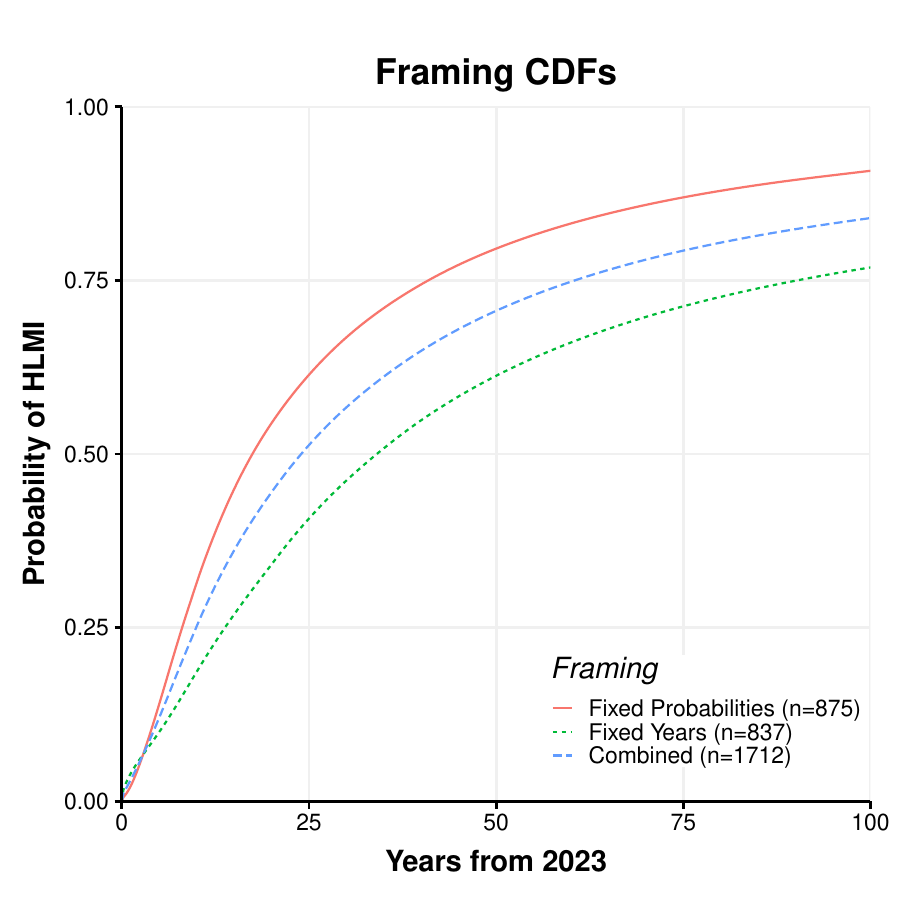}
\caption{\textbf{Participants who received questions framed in terms of fixed-years had later forecasts than those who received questions in terms of fixed-probabilities.}}
\label{fig:hlmi_fixed_years_framing}
\end{figure}
\vfill
\hspace{0pt}
\FloatBarrier

\subsection{How soon will Full Automation of Labor be feasible?}
\begin{table}[h]
\centering
\resizebox{6.5in}{!}{
\label{tab:hlmi_year_by_survey_year}
\begin{tabular}{r|ccc}
& 2016 aggregate forecast & 2022 aggregate forecast & 2023 aggregate forecast \\
\hline
Year with a 50\% chance of HLMI & 2061 (n*=259) & 2060 (n=461) & 2047 (n=1714) \\
Year with a 10\% chance of HLMI & 2025 (n*=259) & 2029 (n=461) & 2027 (n=1714) \\
Year with a 50\% chance of FAOL & 2138 (n*=97) & 2164 (n=202) & 2116 (n=774) \\
Year with a 10\% chance of FAOL & 2036 (n*=97) & 2050 (n=202) & 2037 (n=774) \\
\hline
\end{tabular}}
\caption{\textbf{By what year will human-level performance on all tasks (HLMI) or occupations (FAOL) be feasible?} Aggregate forecasts for HLMI and FAOL in 2023 have gotten earlier since 2022. Comparing 2023 to 2016, 2023’s 50\% estimates are earlier, but the 10\% estimates are later. *n reported for 2016 only is total responses rather than valid responses after cleaning. 
}
\end{table}

\FloatBarrier
\newpage
\subsection{Will there be an intelligence explosion?}
\begin{figure}[h]
\centering
\includegraphics[width=3.2in]{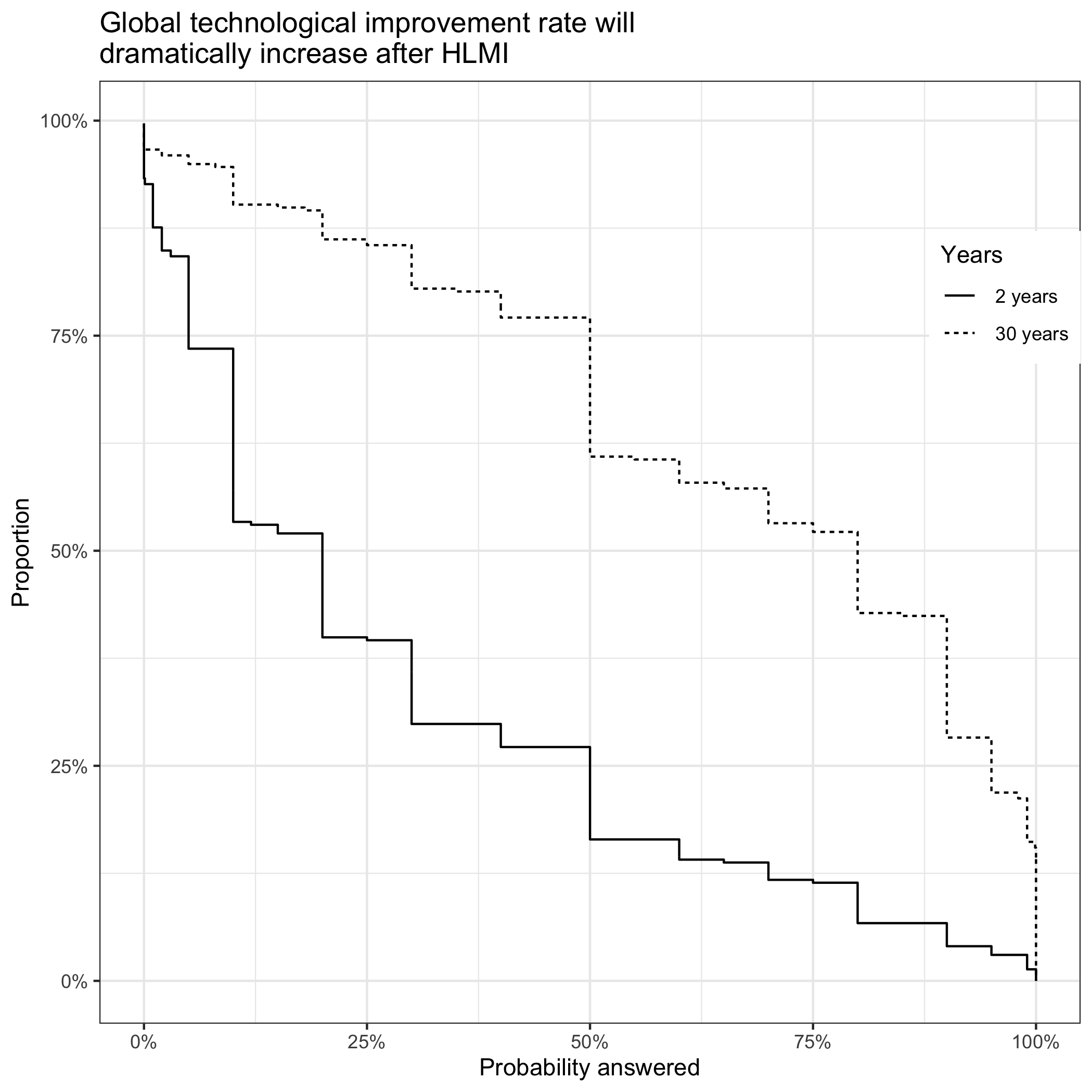}
\includegraphics[width=3.2in]{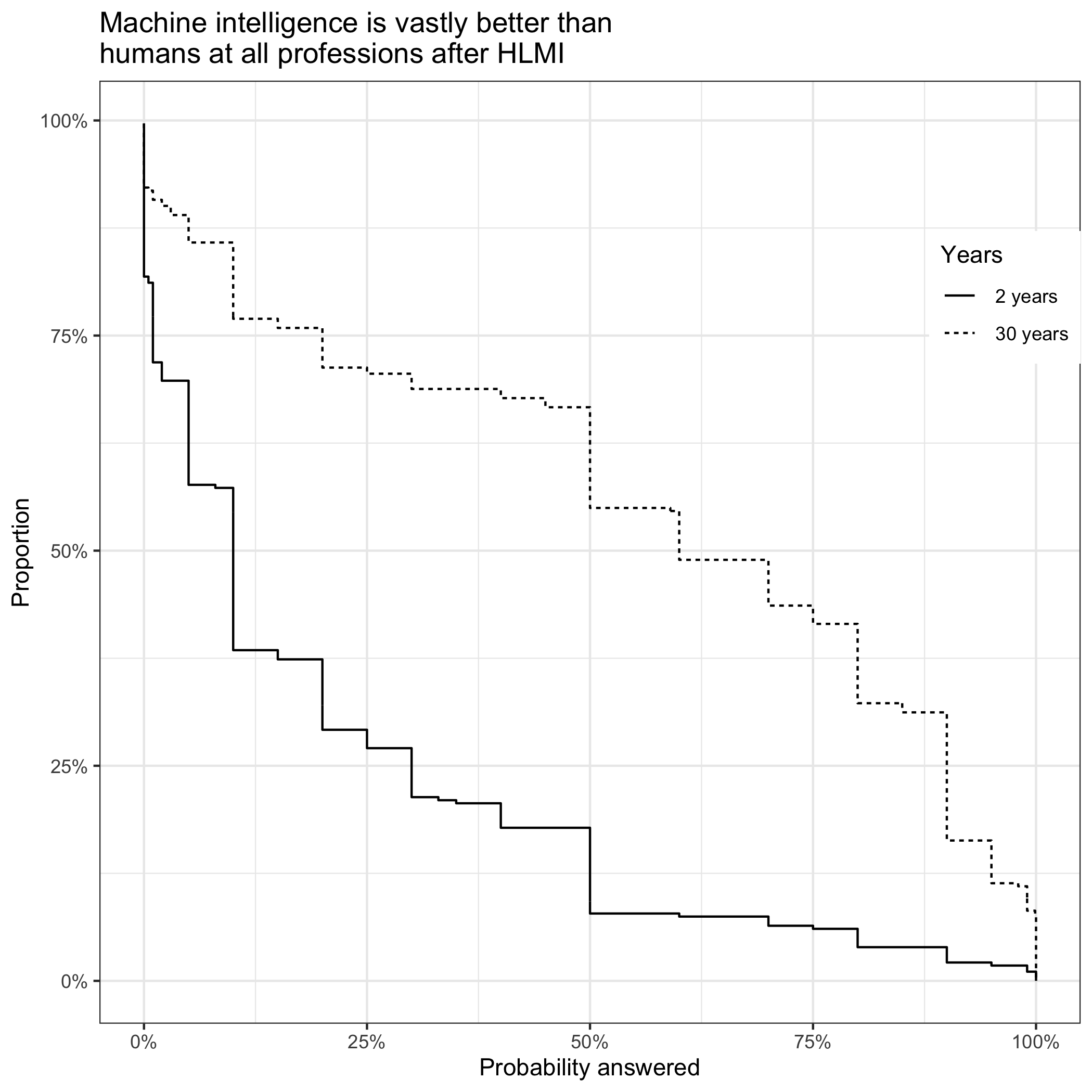}
\label{fig:explosive_improvement}
\label{fig:vastly_better}
\caption{\textbf{Two framings of the intelligence explosion question:} How likely is an ``explosive global technological improvement'' two and 30 years after HLMI? And, How likely is it that AI be “vastly better” than humans in all tasks two and 30 years after HLMI? In the first framing, the median prediction for 2 years post-HLMI was 20\%, whereas the median prediction for 30 years post-HLMI was 80\%. In the second framing, the median prediction for 2 years post-HLMI was 10\%, whereas the median prediction for 30 years post-HLMI was 60\%.}
\end{figure}

\FloatBarrier
\newpage
\subsection{How much should AI safety research be prioritized?}
\begin{figure}[h]
\centering
\includegraphics[width=6in]{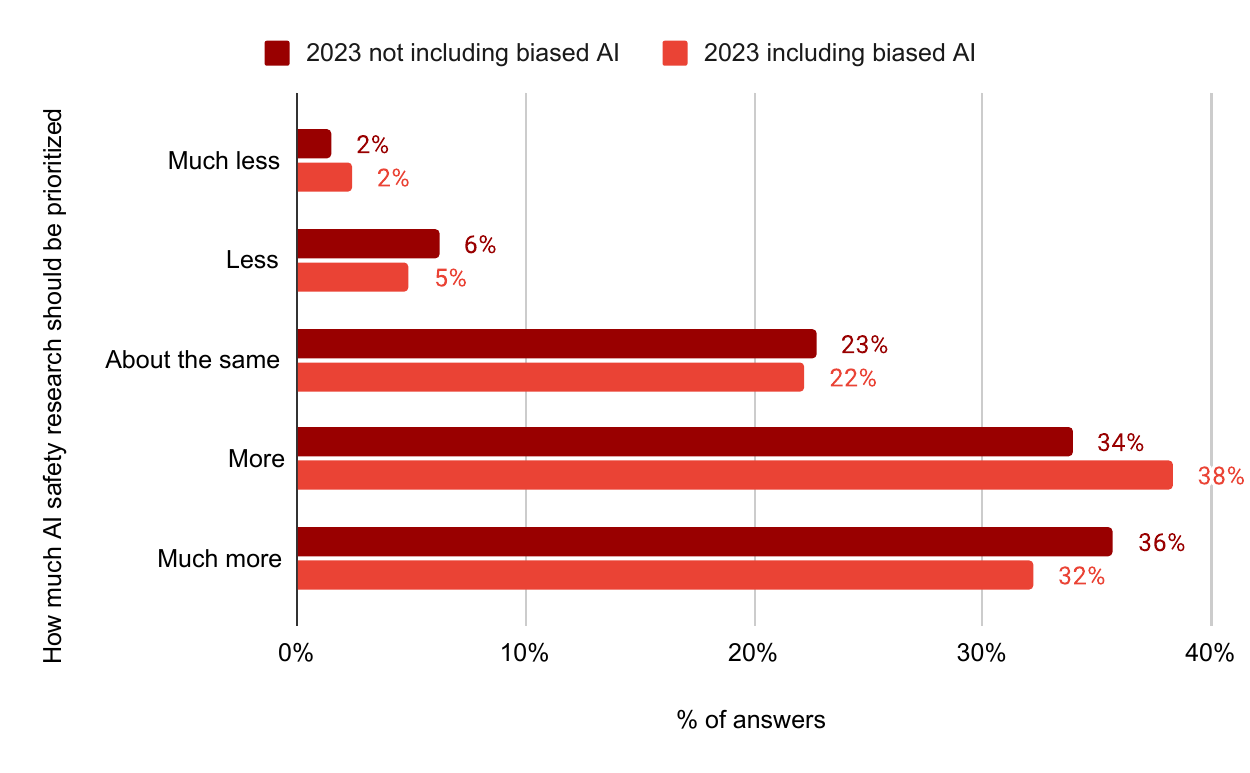}
\label{fig:biased_ai_or_not}
\caption{\textbf{Two framings of the question “How much should AI safety research be prioritized?”, one including and one not including biased AI as an example.}}
\end{figure}
\FloatBarrier
\newpage

\section{Full Descriptions of Milestones} \label{app:milestone_descriptions}
\FloatBarrier

This list is also available at \href{https://tinyurl.com/aitasks}{https://tinyurl.com/aitasks}.

\resizebox{6.5in}{!}{\begin{tabular}{p{2in}p{4.5in}}
Short name & Full description from question \\
\hline
Physically install wiring in a house & Given a one-sentence description of the task and given the same information you would give a human to perform this task (such as information about the house), physically install the electrical wiring in a new home, without more input from humans. \\
Finetune LLM & Given a one-sentence description of the task, download and finetune an existing open source LLM, without more input from humans. The fine-tune must improve the performance of the LLM on some predetermined benchmark metric. \\
Find and patch security flaw & Given a one-sentence description of the task and no more input from humans, find and patch a security flaw in an open source project with over 100,000 users. \\
Run ML study and write paper & Given a one sentence description of a research question in machine learning, conduct a study that would inform the answer to that question and write a paper of a quality that could be accepted at a leading machine learning conference, without more input from humans. \\
Replicate ML paper & Given a study published at a leading machine learning conference, replicate the study without more input from humans. The replication must meet the standards of the ML Reproducibility Challenge (https://paperswithcode.com/rc2022). \\
Build payment processing website & Given a set of specifications, build a website from scratch that can handle payment processing, including the frontend, backend, and secure payment integration, without more input from humans. \\
Long unsolved math problem & Given a list of long-standing unsolved problems in mathematics, such as the Millennium Prize problems or one the problems in ``Unsolved Problems on Mathematics for the 21st Century,'' solve one without more input from humans. \\
Translate text from a newly discovered language using a bilingual document & Translate a text written in a newly discovered language into English as well as a team of human experts, using a single other document in both languages (like a Rosetta stone). Suppose all of the words in the text can be found in the translated document, and that the language is a difficult one. \\
Translate new language speech using films with subtitles and existing bilingual data & Translate speech in a new language given only unlimited films with subtitles in the new language. Suppose the system has access to training data for other languages, of the kind used now (e.g. same text in two languages for many languages and films with subtitles in many languages). \\
Translate text nearly as well as a bilingual but untrained translator & Perform translation about as good as a human who is fluent in both languages but unskilled at translation, for most types of text, and for most popular languages (including languages that are known to be difficult, like Czech, Chinese and Arabic). \\
Offer phone banking services, including unique tasks, on par with human operators & Provide phone banking services as well as human operators can, without annoying customers more than humans. This includes many one-off tasks, such as helping to order a replacement bank card or clarifying how to use part of the bank website to a customer. \\
Classify unseen objects into categories after training on different but similar classes & Correctly group images of previously unseen objects into classes, after training on a similar labeled dataset containing completely different classes. The classes should be similar to the ImageNet classes. \\
\end{tabular}}

\resizebox{6.5in}{!}{\begin{tabular}{p{2in}p{4.5in}}
Recognize a new object in varied settings after seeing it just once & One-shot learning: see only one labeled image of a new object, and then be able to recognize the object in real world scenes, to the extent that a typical human can (i.e. including in a wide variety of settings). For example, see only one image of a platypus, and then be able to recognize platypuses in nature photos. The system may train on labeled images of other objects.

Currently, deep networks often need hundreds of examples in classification tasks\textsuperscript{1}, but there has been work on one-shot learning for both classification\textsuperscript{2}and generative tasks.\textsuperscript{3}

1: Lake et al. (2015). Building Machines That Learn and Think Like People

2: Koch (2015). Siamese Neural Networks for One-Shot Image Recognition

3: Rezende et al. (2016). One-Shot Generalization in Deep Generative Models \\
Create a 3D model and realistic video from a new angle of a scene & See a short video of a scene, and then be able to construct a 3D model of the scene good enough to create a realistic video of the same scene from a substantially different angle.
For example, constructing a short video of walking through a house from a video taking a very different path through the house. \\
Transcribe speech in noise and varied accents on par with humans & Transcribe human speech with a variety of accents in a noisy environment as well as a typical human can. \\
Output written text as a recording indistinguishable from a voice actor & Take a written passage and output a recording that can’t be distinguished from a voice actor, by an expert listener. \\
Prove and generate math theorems publishable in top journals & Routinely and autonomously prove mathematical theorems that are publishable in top mathematics journals today, including generating the theorems to prove. \\
Win Putnam Math Competition (problems with known but very difficult answers) & Perform as well as the best human entrants in the Putnam competition—a math contest whose questions have known solutions, but which are difficult for the best young mathematicians.\\
Beat humans at Go (after same \# games of training) & Defeat the best Go players, training only on as many games as the best Go players have played.

For reference, DeepMind’s AlphaGo has probably played a hundred million games of self-play, while Lee Sedol has probably played 50,000 games in his life.\textsuperscript{1}

1: Lake et al. (2015). Building Machines That Learn and Think Like People \\
Beat ($\geq$ 50\% games) best Starcraft 2 players given only video of screen & Beat the best human Starcraft 2 players at least 50\% of the time, given a video of the screen.

Starcraft 2 is a real time strategy game characterized by:
\begin{itemize}
\item Continuous time play
\item Huge action space
\item Partial observability of enemies
\item Long term strategic play, e.g. preparing for and then hiding surprise attacks. 
\end{itemize}\\
Match human novice in any new computer game in <10 min & Play a randomly selected computer game, including difficult ones, about as well as a human novice, after playing the game less than 10 minutes of game time. The system may train on other games. \\
Outperform humans in new Angry Birds levels & Play new levels of Angry Birds better than the best human players. Angry Birds is a game where players try to efficiently destroy 2D block towers with a catapult. For context, this is the goal of the IJCAI Angry Birds AI competition.\textsuperscript{1}

1: aibirds.org \\
Beat pro testers in all Atari games & Outperform professional game testers on all Atari games using no game-specific knowledge. This includes games like Frostbite, which require planning to achieve sub-goals and initially posed problems for deep Q-networks.\textsuperscript{1, 2}

1: Mnih et al. (2015). Human-level control through deep reinforcement learning

2: Lake et al. (2015). Building Machines That Learn and Think Like People \\
\end{tabular}}

\resizebox{6.5in}{!}{\begin{tabular}{p{2in}p{4.5in}}
Beat novices in 50\% of Atari games after 20 min play & Outperform human novices on 50\% of Atari games after only 20 minutes of training play time and no game specific knowledge.

For context, the original Atari playing deep Q-network outperforms professional game testers on 47\% of games,\textsuperscript{1} but used hundreds of hours of play to train.\textsuperscript{2}

1: Mnih et al. (2015). Human-level control through deep reinforcement learning

2: Lake et al. (2015). Building Machines That Learn and Think Like People \\
Fold laundry as well and as fast as the median human clothing store employee & Fold laundry as well and as fast as the median human clothing store employee. \\
Beat fastest human runners in a 5km city streets race using bipedal robot body & Beat the fastest human runners in a 5 kilometer race through city streets using a bipedal robot body. \\
Build any LEGO set using using non-specialized robotics, with instructions & Physically assemble any LEGO set given the pieces and instructions, using non-specialized robotics hardware.

For context, Fu 2016\textsuperscript{1} successfully joins single large LEGO pieces using model based reinforcement learning and online adaptation.

1: Fu et al. (2016). One-Shot Learning of Manipulation Skills with Online Dynamics Adaptation and Neural Network Priors \\
Efficiently sort large number lists beyond training size & Learn to efficiently sort lists of numbers much larger than in any training set used, the way Neural GPUs can do for addition,\textsuperscript{1} but without being given the form of the solution.

For context, the original Neural Turing Machines could not do this,\textsuperscript{2} but Neural Programmer-Interpreters\textsuperscript{3} have been able to do this by training on stack traces (which contain a lot of information about the form of the solution).

1: Kaiser \& Sutskever (2015). Neural GPUs Learn Algorithms

2: Zaremba \& Sutskever (2015). Reinforcement Learning Neural Turing Machines

3: Reed \& de Freitas (2015). Neural Programmer-Interpreters \\
Write readable Python code for algorithms like quicksort from specs and examples & Write concise, efficient, human-readable Python code to implement simple algorithms like quicksort. That is, the system should write code that sorts a list, rather than just being able to sort lists.

Suppose the system is given only:
\begin{itemize}
\item A specification of what counts as a sorted list
\item Several examples of lists undergoing sorting by quicksort 
\end{itemize}\\
Answer Googleable factoid questions better than expert (w/ web) & Answer any ``easily Googleable'' factoid questions posed in natural language better than an expert on the relevant topic (with internet access), having found the answers on the internet.

Examples of factoid questions:
\begin{itemize}
\item ``What is the poisonous substance in Oleander plants?''
\item ``How many species of lizard can be found in Great Britain?''
\end{itemize}\\
Answer Googleable but open-ended factual questions better than expert (w/ web) & Answer any ``easily Googleable'' factual but open ended question posed in natural language better than an expert on the relevant topic (with internet access), having found the answers on the internet.

Examples of open ended questions:
\begin{itemize}
\item ``What does it mean if my lights dim when I turn on the microwave?''
\item ``When does home insurance cover roof replacement?''
\end{itemize}\\

\end{tabular}}

\resizebox{6.5in}{!}{\begin{tabular}{p{2in}p{4.5in}}
Answer undecided factual questions & Give good answers in natural language to factual questions posed in natural language for which there are no definite correct answers.

For example: ``What causes the demographic transition?'', ``Is the thylacine extinct?'', ``How safe is seeing a chiropractor?'' \\
Write high-grade, unique high school history essays without plagiarizing & Write an essay for a high-school history class that would receive high grades and pass plagiarism detectors.

For example answer a question like ``How did the whaling industry affect the industrial revolution?'' \\
Create songs that can hit the US Top 40 (full audio file) & Compose a song that is good enough to reach the US Top 40. The system should output the complete song as an audio file. \\
Fake new song indistinguishable from a specific artist's work by expert listeners & Produce a song that is indistinguishable from a new song by a particular artist, e.g. a song that experienced listeners can’t distinguish from a new song by Taylor Swift. \\
Write a novel or story that could land on the NYT best-seller list & Write a novel or short story good enough to make it to the New York Times best-seller list. \\
\end{tabular}}

\resizebox{6.5in}{!}{\begin{tabular}{p{2in}p{4.5in}}
Concisely and completely explain AI's computer game moves & For any computer game that can be played well by a machine, explain the machine’s choice of moves in a way that feels concise and complete to a layman. \\
Play poker well enough to win the World Series of Poker & Play poker well enough to win the World Series of Poker. \\
Deduce and symbolize physical laws (e.g. Newtonian mechanics) of a virtual world & After spending time in a virtual world, output the differential equations governing that world in symbolic form.

For example, the agent is placed in a game engine where Newtonian mechanics holds exactly and the agent is then able to conduct experiments with a ball and output Newton’s laws of motion. \\
\end{tabular}}
\newpage

\section{Additional Information on Participation Bias and Question-level Response Bias} \label{app:response_bias}
As with any survey, our results could be skewed by participation bias, if participants had systematically different opinions than those who chose not to participate. Here we review evidence about the presence of participation bias. We find no evidence suggesting strong participation bias.

We sought to minimize participation bias in several ways. First, we made efforts to increase the response rate in ways we expected to attract broad participation not particularly correlation with opinion. For instance, we paid participants \$50 or an equivalent reward for taking the survey (which we estimated to take a median 16 minutes). We also sent a pre-notification email, in it invited questions, and discussed concerns with researchers who wrote. We also sent at least four reminders to take the survey.

As well as aiming to attract a large and broad set of respondents, we tried to limit the ability of recipients to choose to enter the survey based on their opinions, by limiting cues about the survey content before entering the survey. In particular we said the topic was `the future of AI progress' (for example letters, see Appendix 5). Our emails did include links to past ESPAI surveys and included some of our names and affiliations (University of Oxford, AI Impacts, University of Bonn), so evidence about the topic and the authors was available to participants who investigated or were previously familiar with us or the survey (increasingly likely, since our past surveys received substantial public attention, and have now been taken before by many researchers). 

As well as bias in who chooses to take a survey, there can be bias in who chooses to answer each particular question within it. All but one question could be skipped. However each question in this survey was answered by on average 96\% of those who saw it, excluding demographics questions, free response questions, and questions asking for a response conditionally. So the scope for question-level participation bias in opinion questions reported on in this paper is very small.

The question which was skipped most often (answered by 90.3\% of those who saw it) was about the number of years until the occupation of ``AI researcher'' would be fully automatable, with the fixed-probabilities framing. The question which was skipped the least often was about the long-term value of HLMI (see Section~\ref{sec:goodness_of_hlmi}), answered by 100\% of those who saw it, probably because it was not possible to skip and continue the survey.

For random samples of responders (n=369) and of non-responders (n=589), we compared gender, region, PhD start year, number of citations, and work in industry or academia, based on our best guesses from available public data. The most substantial difference between the groups was that those who worked in industry were 61\% as likely to respond as the base rate. Women were 66\% as likely to respond as the base rate, though were a relatively small portion of respondents or non-respondents overall (around one in ten), which limits how much this could affect the aggregate results. Those with at least \numprint{1000} citations were 69\% as likely to respond as the base rate, and people in Asia were 84\% as likely as the base rate to respond. We checked for correlations between the collected demographic data and respondents' beliefs about the arrival or overall outcome of HLMI. Most demographic factors showed minimal influence, but female respondents generally expected less extreme positive or negative outcomes. Additionally, respondents whose undergraduate education was in Asia anticipated an 11 year earlier arrival of High-Level Machine Intelligence (HLMI) than participants from Europe, North America, or other regions combined. See Appendix~\ref{app:demographics} for more demographic comparisons.

Another source of evidence comes from people’s reported interest in the topic. One particular concern might be that researchers who have a strong interest in the future of AI might be more likely to participate than those who do not. However, when asked ``How much thought have you given in the past to when HLMI (or something similar) will be developed?'' only 7.6\% said ``A great deal.'' And when asked ``How much thought have you given in the past to social impacts of smarter-than-human machines?'' only 10.3\% said ``A great deal.'' This would seem to rule out the possibility that a large proportion of respondents were people for whom the future of AI is an ideological issue.

\section{Links to Data} \label{app:data_links}
\href{https://tinyurl.com/espai2023-clean-anon-responses}{Anonymized and cleaned responses: https://tinyurl.com/espai2023-clean-anon-responses}

\href{https://tinyurl.com/espai2023-invite}{Example invitation letter: https://tinyurl.com/espai2023-invite}

\href{https://tinyurl.com/aitasks}{List of task descriptions: https://tinyurl.com/aitasks}

\href{https://tinyurl.com/espai2023-response-counts}{The number of participants who saw and answered each question: https://tinyurl.com/espai2023-response-counts}

\href{https://tinyurl.com/espai2023-addenda}{Further data and analysis as it becomes available: https://tinyurl.com/espai2023-addenda}

\section{Yuen's Test (Bootstrap Version) Results} \label{app:yuen}
\resizebox{6in}{!}{%
\begin{tabular}{r|ccccc}
Variable & P-value & T$_y$ & Trimmed mean difference & Lower bound CI & Upper bound CI \\
\hline
hlmi\_50percent & \cellcolor{green!25}0 & 4.96 & 12.49 & 7.58 & 17.39 \\
faol\_50percent & \cellcolor{green!25}0.0052 & 2.84 & 37.54 & 13.08 & 62 \\
task\_Rosetta\_50percent & 0.1106 & 1.62 & 2.52 & -0.66 & 5.7 \\
task\_translatespeech\_50percent & \cellcolor{green!25}0.0072 & 2.74 & 2.7 & 0.77 & 4.63 \\
task\_translatetext\_50percent & 0.258 & 1.25 & 2.69 & -2.25 & 7.62 \\
task\_phonebanking\_50percent & 0.1862 & 1.28 & 1.23 & -0.67 & 3.12 \\
task\_groupimages\_50percent & 0.8058 & 0.24 & 0.23 & -1.79 & 2.25 \\
task\_oneshotlearning\_50percent & 0.4778 & 0.72 & 1.2 & -2.25 & 4.65 \\
task\_3D\_50percent & \cellcolor{green!25}0.0366 & 2.09 & 1.9 & 0.15 & 3.65 \\
task\_transcribe\_50percent & \cellcolor{green!25}0.0116 & 3.29 & 2.74 & 1.06 & 4.41 \\
task\_recording\_50percent & \cellcolor{yellow!25}0.0748 & 2.1 & 1.92 & -0.16 & 3.99 \\
task\_provemath\_50percent & 0.9864 & -0.02 & -0.08 & -7.36 & 7.2 \\
task\_Putnam\_50percent & 0.3896 & 0.91 & 2.92 & -4.94 & 10.77 \\
task\_Go\_50percent & 0.3376 & 0.95 & 1.55 & -1.59 & 4.68 \\
task\_Starcraft\_50percent & 0.5176 & -0.63 & -0.45 & -1.81 & 0.92 \\
task\_randomPCgame\_50percent & 0.21 & 1.41 & 2.4 & -1.49 & 6.3 \\
task\_AngryBirds\_50percent & 0.5688 & 0.55 & 0.22 & -0.64 & 1.08 \\
task\_profAtari\_50percent & 0.6984 & -0.39 & -0.44 & -2.67 & 1.8 \\
task\_novAtari\_50percent & 0.1042 & 1.61 & 1.6 & -0.37 & 3.56 \\
task\_laundry\_50percent & 0.7116 & 0.35 & 0.67 & -3.36 & 4.69 \\
task\_run\_50percent & \cellcolor{green!25}0.0362 & 2.19 & 3.54 & 0.29 & 6.8 \\
task\_LEGO\_50percent & 0.6892 & -0.4 & -0.44 & -2.6 & 1.72 \\
task\_sortlist\_50percent & 0.5616 & 0.58 & 0.85 & -2.06 & 3.75 \\
task\_Python\_50percent & \cellcolor{yellow!25}0.0768 & 1.95 & 2.52 & -0.65 & 5.69 \\
task\_Googleable\_50percent & 0.1606 & 1.55 & 1.6 & -1.09 & 4.29 \\
task\_openGoogleable\_50percent & \cellcolor{green!25}0.0398 & 2.38 & 2.25 & 0.19 & 4.32 \\
task\_NoDefAnswer\_50percent & \cellcolor{green!25}0.0258 & 2.49 & 5 & 1.11 & 8.9 \\
task\_essay\_50percent & \cellcolor{yellow!25}0.0912 & 1.69 & 1.24 & -0.22 & 2.69 \\
task\_songTop40\_50percent & 0.1654 & 1.41 & 2.64 & -1.24 & 6.52 \\
task\_songartist\_50percent & \cellcolor{yellow!25}0.0958 & 1.57 & 2.38 & -0.45 & 5.22 \\
task\_bestseller\_50percent & \cellcolor{green!25}0.0038 & 3.59 & 6.06 & 2.42 & 9.7 \\
task\_explainmove\_50percent & 0.2466 & 1.17 & 1.76 & -1.28 & 4.79 \\
task\_poker\_50percent & 0.3718 & 0.97 & 1.04 & -1.44 & 3.52 \\
task\_virtualworld\_50percent & \cellcolor{yellow!25}0.0578 & -1.85 & -3.53 & -7.19 & 0.12 \\
truckdriver\_50percent & 0.908 & -0.12 & -0.14 & -2.68 & 2.4 \\
surgeon\_50percent & \cellcolor{green!25}0.0408 & 2.06 & 5.92 & 0.29 & 11.55 \\
retail\_salesperson\_50percent & \cellcolor{green!25}0.006 & 2.76 & 3.48 & 1.01 & 5.94 \\
AIresearcher\_50percent & \cellcolor{green!25}0.0066 & 3.72 & 21.84 & 8.63 & 35.06 \\
final\_occupation\_50percent & \cellcolor{green!25}0.0264 & 2.59 & 31.73 & 4.52 & 58.95 \\
\hline
\end{tabular}}

\end{document}